\documentclass[twocolumn,superscriptaddress,amsmath,amssymb,aps,prb,floatfix]{revtex4-2}

\usepackage{graphicx}
\usepackage{bm}
\usepackage{physics}
\usepackage[colorlinks,
    linkcolor=blue,
    anchorcolor=red,
    citecolor=green
]{hyperref}
\usepackage{wasysym}
\usepackage{MnSymbol}
\usepackage{xcolor}

\newcommand{\PRLSection}[1]{{\color{blue} \textit{ #1}}}

\usepackage{wasysym}

\renewcommand{\v}[1]{\mathbf{#1}}
\newcommand{\arctanh}{\text{artanh}}

\begin{document}
\title{Topological Green's function zeros in an exactly solved model and beyond}
\author{Steffen Bollmann}
\affiliation{%
Max-Planck Institute for Solid State Research, 70569 Stuttgart, Germany
}%
\author{Chandan Setty}
\affiliation{%
Department of Physics and Astronomy, Rice Center for Quantum Materials, Rice University, Houston, Texas 77005, USA
}%
\author{Urban F. P. Seifert}
\affiliation{%
Kavli Institute for Theoretical Physics, University of California, Santa Barbara, CA 93106, USA
}%

\author{Elio J. K\"onig}
\affiliation{%
Max-Planck Institute for Solid State Research, 70569 Stuttgart, Germany
}%

\begin{abstract}
The interplay of topological electronic band structures and strong interparticle interactions provides a promising path towards the constructive design of robust, long-range entangled many-body systems.
 As a prototype for such systems, we here study an exactly integrable, local model for a fractionalized topological insulator. Using a controlled perturbation theory about this limit, we demonstrate the existence of topological bands of zeros in the exact fermionic Green's function and show that {in this model} they do affect the topological invariant of the system, but not the quantized transport response. 
 Close to (but prior to) the Higgs transition signaling the breakdown of fractionalization, the topological bands of zeros acquire a finite ``lifetime''. We also discuss the appearance of edge states and edge zeros at real space domain walls separating different phases of the system. 
 This model provides a fertile ground for controlled studies of the phenomenology of Green's function zeros and the underlying exactly solvable lattice gauge theory illustrates the synergetic cross-pollination between solid-state theory, high-energy physics and quantum information science.
\end{abstract}

\maketitle

The spectrum of low-lying excitations is central to quantum-many body physics and reflected in poles of the electronic Green's function (GF)~\cite{AGD}. Interpreted as a propagator, the GF encodes quantum mechanical transition amplitudes. From this perspective, it is a natural and tantalizing question to also {study} nodes of the amplitudes {indicating} 
destructive many-body interference. 
{This results in GF zeros which are key to Fermi surface reconstruction and pseudogap physics --- a major open problem in quantum materials research~\cite{Dzyaloshinskii2003,SakaiImada2016,GazitSachdev2020,KoenigTsvelik2020,Fabrizio2022,Fabrizio2023,VasiliouBultinck2023,NikolaenkoSachdev2023}.} 
Just like their counterpart, the poles, GF zeros defined by $\mathbf G^R(\epsilon_\emptyset(\v k), \v k) \psi(\v k)= 0$
trace a bandstructure $\epsilon_\emptyset(\v k)$ in the Brillouin zone (BZ). The associated ``Bloch''-like eigenstates $\psi(\v k)$ may be topological~\cite{Gurarie2011,SettySi2023a,SettySi2023b, SettySi2023c, MaiPhillips2023, ManningBradlyn2023} leading to, e.g., topological boundary zeros~\cite{WagnerSangiovanni2023}.
{Beyond numerics, analytically exact results for \textit{Hatsugai-Kohmoto-type} models with ultra-non-local interactions\cite{HatsugaiKohmoto1992, Phillips2018, Setty2021a, Setty2021b, PhillipsHuang2020} demonstrate non-trivial band representations of poles and zeros, alike\cite{SettySi2023a, SettySi2023c, MaiPhillips2023, JablonskiWisokinski2023}.}

\begin{figure}
    \centering
    \includegraphics[width=.45\textwidth]{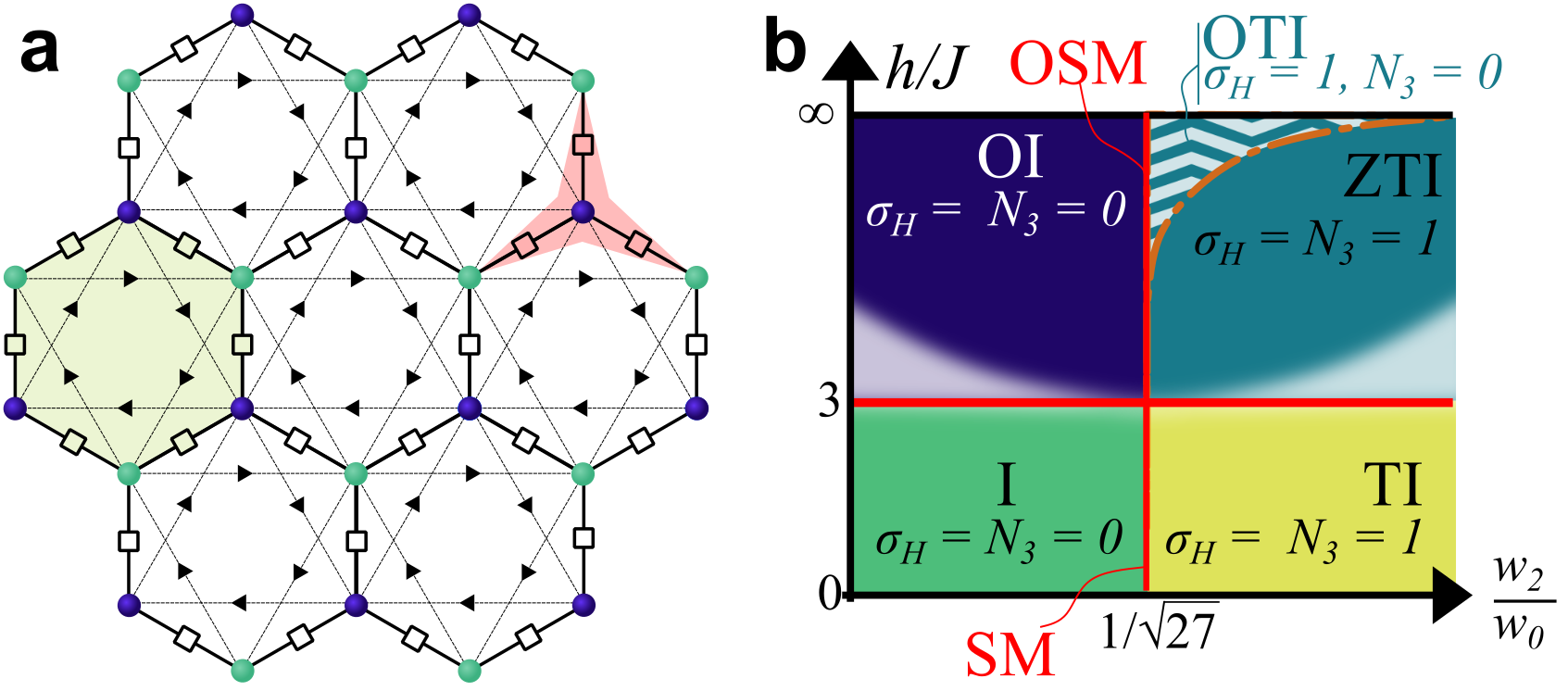}
    \caption{a) Schematics of the model. b) Phase diagram as a function of $h/J$ and $w_2/w_0$. Acronyms: I = insulator, TI = topological insulator, OTI = orthogonal TI, ZTI = zeros TI, SM = semimetal, OSM = orthogonal SM. In the OTI phase, the poles of the GF are topological, but the zeros of the GF are trivial. In the ZTI phase, both are topological. The shaded region close to the horizontal confinement-deconfinement transition at $h/J =3$ indicates that GF zeros acquire a ``lifetime''.} 
    \label{fig:Schematics}
\end{figure}

In this letter, we study the topological properties of Bloch-bands of zeros by investigating a local many-body Hamiltonian which is exactly soluble,
and in which fermions couple to an emergent $\mathbb{Z}_2$ gauge theory \cite{NandkishoreSenthil2012,ZhongLuo2012,ZhongLuo2013,MaciejkoRuegg2013,GazitVishwanath2017,GazitSachdev2020,KoenigTsvelik2020,BorlaMoroz2022,RoyKoenig2023,GoltermanShamir2023,EmontsZohar2023}.
In this theory, correlators of physical fermions are given by convolutions of propagators of fractionalized excitations. Equipped with this fundamental structure we can exactly prove in the soluble limit (i.e.~with a static $\mathbb{Z}_2$ gauge field) that the physical GFs possess zeros. This provides us with a starting point for 
a controlled perturbative expansion about the integrable limit, in which we demonstrate that dispersing GF zeros form a topologically non-trivial bandstructure, but that their existence is not directly related to quantized topological transport coefficients.
An infinite order resummation of leading Feynman diagrams allows to access the exact long-wavelength GF in the vicinity of the deconfinement-confinement Higgs transition.
As the transition is approached, the mass of GF zeros increases and hits the continuum of excitations (i.e. poles) in the system. Beyond this point, the energy of GF zeros becomes complex. 

A GF based topological invariant~\cite{IshikawaMatsuyama1986,volovikYakovenko1989,Volovik2003,Gurarie2011,EssinGurarie2011,RaghuZhang2008} may be employed to attribute topology to strongly interacting systems, e.g., for Altland-Zirnbauer class A in two dimensions
 \begin{equation}
N_3[\mathbf G] =  \int \frac{\text d \epsilon \text d^2 k}{24\pi^2} \epsilon^{\mu \nu \rho} \tr [\mathcal A_\mu \mathcal A_\nu\mathcal A_\rho],
\label{eq:Volovik}
\end{equation}
where $\mathcal A_\mu = \mathbf G^{-1}(i \epsilon, \v k)\partial_{K_\mu} \mathbf G(i\epsilon, \v k)$. In the non-interacting limit $N_3$ is identified with the quantized anomalous Hall conductance $\sigma_H$, but the equivalence may break down in the presence of strong interactions~\cite{ZhaoPhillips2023, SettySi2023b, BlasonFabrizio2023,GavenskyGoldman2023}. Note that GF zeros enter $N_3$ analogously to poles~\cite{Gurarie2011} and topological transitions in $N_3$ may occur through gap closing of either poles or zeros. 
Related to this, the free Dirac fermion at the non-interacting topological transition may experience symmetric mass generation~\cite{ManmanaGurarie2012,SlagleXu2015,You-Vishwanath2018,WangYou2022} at strong coupling to account for the topological transition in the GF invariant without requiring a gap closure in the physical fermionic spectrum.

\PRLSection{Model.} In this letter, we consider a system in which Haldane's model for the quantum Hall effect without Landau levels is coupled to the toric code (TC) on a honeycomb lattice, cf. Fig.~\ref{fig:Schematics} a), $H  =H_{\rm QAH} +H_{\rm TC}+  H_J$ where~\cite{Haldane1988,ZhongLuo2013} 

\begin{subequations}
\begin{align}
    H_{\rm QAH}  = &- w_0 \sum_{\v x} (-1)^{\v x} c^\dagger_{\v x} c_{\v x} - w_1 \sum_{\langle \v x, \v x' \rangle}  c^\dagger_{\v x} Z_{\langle \v x, \v x' \rangle} c_{\v x'}  \notag \\
    &- i w_2 \sum_{\llangle \v x, \v x' \rrangle} (-1)^{\v x} [c^\dagger_{\v x} \prod _{ \v  b\in \gamma_{\v x, \v x'} }Z_{\v b} c_{\v x'}- H.c.  ] \label{eq:Hfermions},\\
    H_{\rm TC} =& - K_{\varhexagon} \sum_{\varhexagon} B_{\varhexagon} - h \sum_{\v x} Q_{\v x},\\
    H_J  =& -J \sum_{\v b} Z_{\v b}.
\end{align}
\label{eq:Htot}
\end{subequations}
Throughout the letter, $X_{\v b},Z_{\v b}$ denote Pauli matrices residing on bonds and $B_{\varhexagon} = \prod_{\v b \in \partial \varhexagon} Z_{\v b}, 
Q_{\v x} = (-1)^{\hat n_{\v x}}\prod_{\v b \in \partial \v x} X_{\v b}$ are plaquette and star operators respectively. We consider $h>0$ and, given that $m-$particles are entirely irrelevant for the present discussion, send $K_{\varhexagon} \rightarrow \infty$.
Note that we added a staggered sublattice potential $w_0$, which contrary to the Haldane next-nearest neighbor hopping $w_2$ opens a trivial gap near the K, K' points. Both $w_0, w_2$ have opposite contributions in opposite sublattices, as indicated by the notation $(-1)^{\v x}$. The path $\gamma_{\v x, \v x'}$ is the shortest connection of next-nearest neighbors. Planck's constant and the speed of light are set to unity throughout this letter.

\begin{figure}
    \centering
    \includegraphics[width=.45\textwidth]{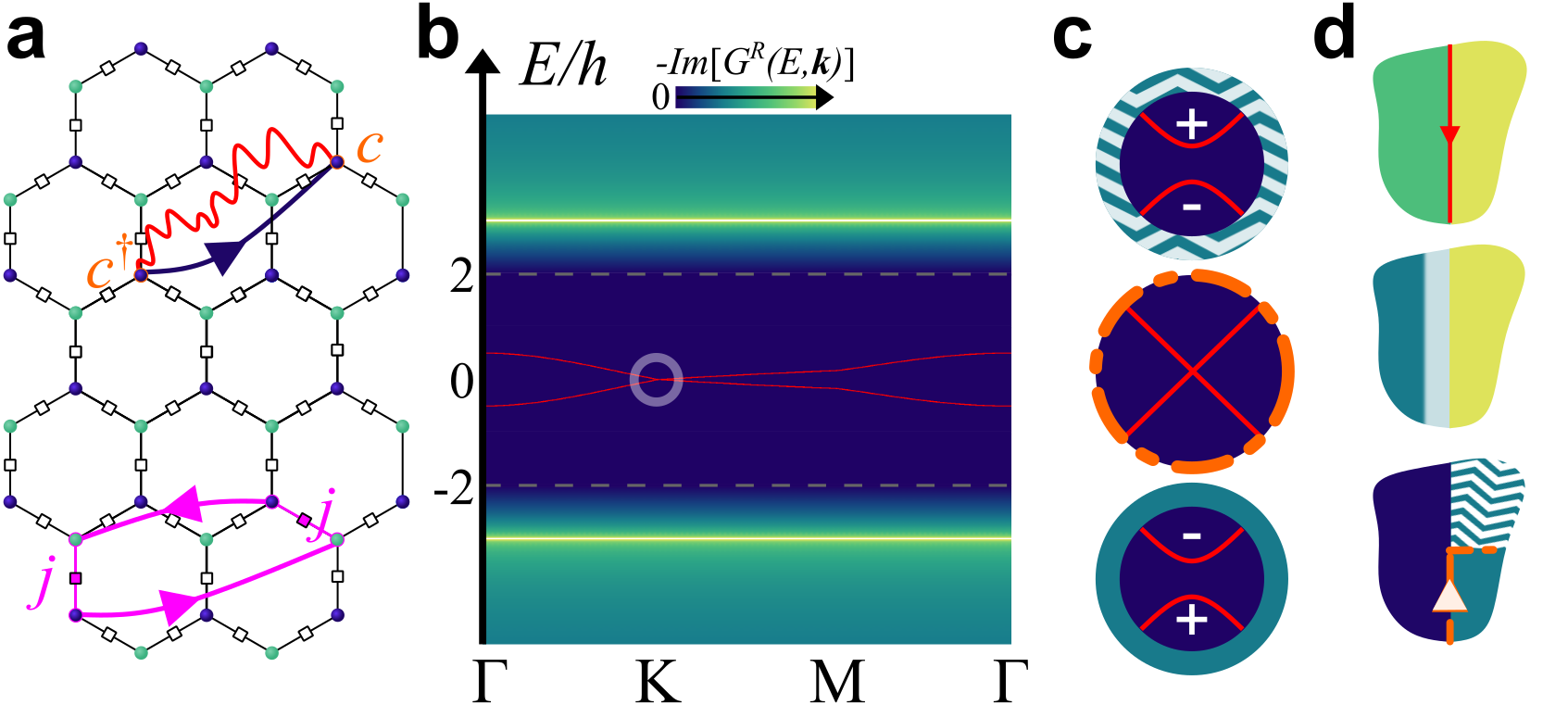}
    \caption{a) Real space representation of the GF, Eq.~\eqref{eq:GF}, and the current-current correlator. In the deconfining phase, the gapped $e$-particles (red, wiggly propagator) imply an exponentially decaying fermionic GF even when the propagator of orthogonal fermions is gapless. However, $e$-particle propagators do not enter the current-current correlator (pink). b) In the limit $J = 0$ the spectral function is exactly momentum independent (color plot for $h = w_1$ and $w_0 = w_2 = 0$ along a BZ cut). 
Perturbative inclusion of $J$ leads to moderate dispersion of Hubbard bands (not shown), while GF zeros acquire a graphene-like dispersion of bandwidth $\mathcal O(J w_1/h)$ (red). c) The additional inclusion of the staggered potential leads to a trivial gap $\mathcal O(w_0)$ in the dispersion of zeros (top circle). In contrast, Haldane-like next-nearest neighbor hopping opens a topological gap $\mathcal O(J^2 w_2/h^2)$ in the Dirac nodes of the zeros (bottom circle). At $w_0 \sim J^2 w_2/h^2$ the gap vanishes in one of the K/K'-valleys signaling a topological phase transition of GF zeros (central circle). d) Real space diagram of edge states/zeros with color coding of phases as in Fig.~\ref{fig:Schematics} b).  } 
    \label{fig:GreensFunction}
\end{figure}

\PRLSection{Exact solutions.} This model is exactly soluble~\cite{ZhongLuo2013} in two limits. First, when $h/J = 0$, all qubits of the TC are frozen and the ground state is
\begin{subequations}
\begin{equation}
        \ket{GS}_{h/J = 0} = \ket{+}_{TC} \otimes \ket{FS}_c. \label{eq:ProductState}
    \end{equation}
The state $\ket{+}_{TC}$ indicates that $\langle Z_{\v b} \rangle = 1 \forall \v b$ and $\ket{FS}_c$ is the filled Fermi sea at half filling of the free fermion model defined by the Bloch Hamiltonian in sublattice space
\begin{equation}
h(\v k) = \left (\begin{array}{cc}
w_0 + w_2 m(\v k) & -w_1 s(\v k) \\ 
-w_1 s^*(\v k) & -w_0 - w_2 m(\v k)
\end{array}  \right), \label{eq:HBloch}
\end{equation}
where $s(\v k) = 1+2 e^{-i{\sqrt{3} k_y}/{2}} \cos \left({k_x}/{2}\right), m(\v k) = 4 \sin \left({k_x}/{2}\right) [\cos \left({k_x}/{2}\right)-\cos \left({\sqrt{3} k_y}/{2}\right)]$. 
The Hamiltonian Eq.~\eqref{eq:HBloch} is topological (topologically trivial) for $w_2/w_0 >1/\sqrt{27}$ ($w_2/w_0 <1/\sqrt{27}$). We denote this phase as TI (I) in Fig.~\ref{fig:Schematics} b), it clearly displays $\sigma_H = N_3 = 1$ ($\sigma_H = N_3 = 0$) and an edge state at the TI-I interface, Fig.~\ref{fig:GreensFunction} d).

By virtue of the fact that all $Q_{\v x}$ and $B_{\varhexagon}$ are integrals of motion when $J = 0$, Eq.~\eqref{eq:Htot} is also exactly soluble in the limit $h/J \rightarrow \infty$, in this case
    \begin{equation}
        \ket{GS}_{h/J \rightarrow \infty} = \prod_{\v x}\frac{1 + Q_{\v x}}{2} \Big (\ket{+}_{TC} \otimes \ket{FS}_c \Big ). \label{eq:TCGroundstate}
    \end{equation}
\end{subequations}

It is important to emphasize that
Eq.~\eqref{eq:ProductState} is the direct product of a product state in the TC sector and a Slater determinant of a fully filled band. Therefore it is short range entangled and strongly distinct from~\eqref{eq:TCGroundstate} which involves topologically order stemming from the TC sector.

\PRLSection{Propagators of fractionalized theory: $h \gg J$.} The TC coupled to fermions, Eq.~\eqref{eq:Htot}, is related to $\mathbb Z_2$ slave spin theories~\cite{DeMediciBiermann2005,RueggSigrist2010} of Hubbard-like models by gauge fixing \cite{TupitsynKitaev2010}. In this context $h$ is the local Hubbard repulsion, while $w_{1,2}$ and $J$ correspond to the hopping amplitudes of fractionalized fermionic and slave spin excitations, respectively. In particular, the gapped $e$-particles of the TC are given by flipped star operators (i.e. sites at which $Q_{\v x} = -1$ instead of $Q_{\v x} = 1$).
In the integrable limit $J = 0$ the Euclidean time GF $ D(\tau;\v x, \v x') = e^{- 2 h \vert \tau \vert} \delta_{\v x, \v x'}$ of $e$-particles is ultralocal, this can be interpreted as a vanishing kinetic energy of slave-spin-flips. 
The fermionic excitations on top of the filled Fermi sea $\ket{FS}_c$ with propagator $G(i \epsilon, \v k) = [i \epsilon - h(\v k)]^{-1}$ will be referred to as ``orthogonal fermions''. 

As the creation of the physical fermion in the fractionalized theory involves the simultaneous excitation in the fermionic and TC (slave spin) sectors alike, the total two-point function takes the form of a convolution, Fig.~\ref{fig:GreensFunction} a):
\begin{equation}
    \mathbf G(\tau;\v x, \v x') \equiv - \langle \mathcal T [c_{\v x}(\tau)c_{\v x'}^\dagger(0)]\rangle =  D(\tau;\v x, \v x') G(\tau;\v x,\v x'). \label{eq:GF}
\end{equation}
We now go away from the solvable limit $J=0$ and use perturbation theory in $J/h$ to obtain $D$ and $\mathbf G$. 
The perturbative evaluation~\cite{Suppmat} of 
the fermionic propagator up to second order in $J/h$ is conveniently written in the momentum- and frequency domain as
$\mathbf G( i\epsilon,\v k) = - \left [i \epsilon +  \mathcal C_0 h_\emptyset(i\epsilon, \v k) h/  w_1 \right]/{ 4h^2},$ where
\begin{align}
   h_\emptyset(i\epsilon, \v k) & = \left (\begin{array}{cc}
w_0 + w_2 \frac{J^2}{h^2}\mathcal C_2 m(\v k) & - w_1 \frac{J}{h}\mathcal C_1 s(\v k) \\ 
-w_1\frac{J}{h} \mathcal C_1 s^*(\v k) & -w_0 - w_2 \frac{J^2}{h^2} \mathcal C_2 m(\v k)
\end{array}  \right), \label{eq:HZero} 
\end{align}
and the matrix form results from the sublattice-resolved structure.
In the limit $w_{0} \ll w_2 \ll w_1 \ll h, \epsilon \ll h$, the functions $\mathcal C_{0,1,2}$ acquire a finite, positive value of order unity. 

After analytical continuation to real energies, the following observations are noteworthy: First, for $J = 0$, the ultralocal, gapped behavior of $D(\tau; \v x, \v x')$ implies a $\v k$ independent spectral weight, Fig.~\ref{fig:GreensFunction} b), with all spectral weight pushed to ``Hubbard'' bands at energies larger than $2h$. Second, inside the gap there are flat bands of zeros at $E = \pm w_0$ (not shown)\cite{RoyKoenig2023}. One may readily check that in the limit $h/J \rightarrow \infty$, $N_3 = 0$. While this behavior persists throughout, for $w_2/w_0 < 1/\sqrt{27}$ ($w_2/w_0 > 1/\sqrt{27}$) i.e. when $h(\v k)$ is topologically trivial (topological), we use the name orthogonal (topological) insulator, abbreviated as OI (OTI) in Fig.~\ref{fig:Schematics} b).
Third, perturbation theory in $J$ leads to slow $\v k$ dependence of ``Hubbard'' bands (not shown) and importantly lifts the degeneracy of the flat bands of zeros, see red contours in Fig.~\ref{fig:GreensFunction} b). Note that nearest neighbor hopping, $s(\v k)$, enters $\mathbf G(i \epsilon, \v k)$ to first order, while the Haldane-like next-nearest neighbor hopping, $m(\v k)$, appears in the GF only at the second order in $J/h$. Fourth, by comparing Haldane mass and staggered potential of $h_\emptyset$, the band structure of zeros is topological for $h/J \lesssim \sqrt{w_2/w_0}$, Fig.~\ref{fig:GreensFunction} c). We call this phase ``zeros topological insulator'' (ZTI) and explicitly demonstrate below that $N_3 =1$ in the ZTI, just as in the TI phase. This equality follows from $h_\emptyset(\v k)$ entering the numerator of $\mathbf G (i \epsilon, \v k)$ with the opposite sign than $h(\v k)$ in the denominator of $G(i \epsilon, \v k)$. This also implies that the Bloch wave functions $\psi (\v k)$ associated to zeros are not wannierizable, i.e. do not correspond to elementary band representations. 
  Sixth, the topological phase transition between ZTI and OTI involves a Dirac-like gap-closing of zeros, central panel of Fig.~\ref{fig:GreensFunction} c), which is depicted by a dot-dashed orange line in Fig.~\ref{fig:Schematics} b). Finally, in the limit of slowly spatially varying $w_2(\v x)$, semiclassical arguments imply boundary zeros in real space at real space domain walls between OTI and ZTI, Fig.~\ref{fig:GreensFunction} d), bottom panel.

\begin{figure}
    \centering
    \includegraphics[width=.45\textwidth]{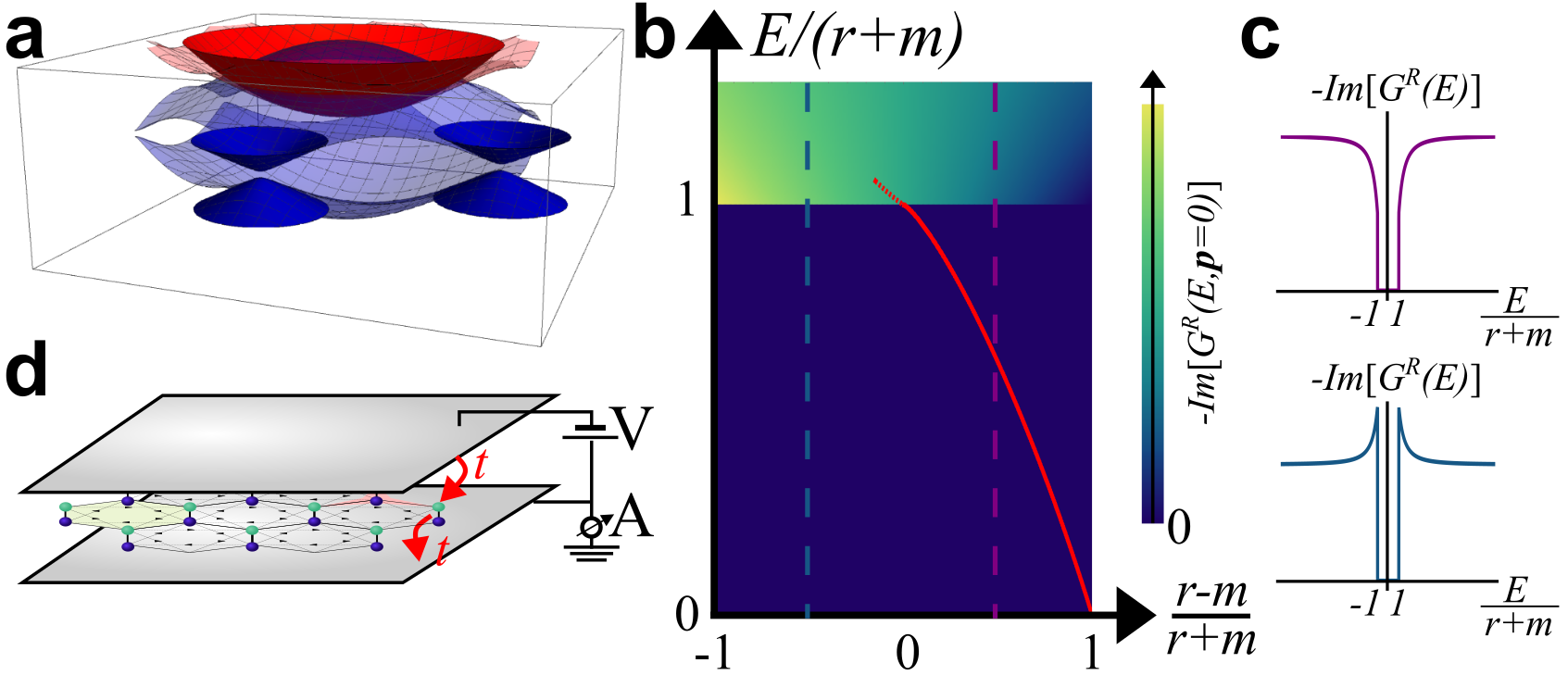}
    \caption{a) Dispersion of low-lying $e$-particles (red), approximated by a hyperbolic dispersion relation, and of low-lying fermions (blue) with effective Hamiltonian Eq.~\eqref{eq:HCont}. b) Color plot: Spectral weight obtained from Eq.~\eqref{eq:ExactGFtotal}. The equation $\det[\mathbf G^R(E,\v p = 0)] = 0$ defines the lowest energy of a GF zero (red). Importantly, this energy acquires an imaginary part (a ``lifetime'') once the solution hits the continuum (i.e. once $r$ drops below $m$). c) Top: (Bottom:) Spectral weight at $r - m = [r+m]/2$ ($r - m = -[r+m]/2$). d) Planar tunneling setup in which the ZTI system is sandwiched between two metallic leads.} 
    \label{fig:ContLimit}
\end{figure}

\PRLSection{Exact continuum GF for $h\gg \sqrt{h^2-3hJ} \gg J^3/h^2$.}
The inclusion of $J$-induced $e$-particle hopping can be resummed to yield dispersive $e$-particle bands, of which the lower can be approximated by $\omega(\v q) = \sqrt{r^2 + v^2 q^2}$ at lowest energies $r \ll h$, Fig.~\ref{fig:ContLimit} a). Here $v = \sqrt{hJ}$, $r^2 = 4h [h - 3J]$. The closing of the gap indicates condensation of $e$-particles, i.e.~the Higgs transition. Below the transition, the fermionic GF, Eq.~\eqref{eq:GF}, acquires a finite quasiparticle weight~\cite{GazitSachdev2020,KoenigTsvelik2020}.

In the present case of 2+1 dimension, the immediate vicinity of the transition is governed by non-Gaussian fluctuations ($e$-particle self-interaction), but extending standard arguments by Ginzburg and Levanyuk~\cite{Levanyuk1959,Ginzburg1960,Suppmat}, one may neglect those as long as $r\gg J^3/h^2$~\cite{Suppmat,KoenigTsvelik2020}. 
Moreover, in the present model, the interaction between fermion and $e$-particles is absent to leading order~\cite{KoenigTsvelik2020}. Such terms might be generated due to fluctuations near the confinement transition, but are renormalization group irrelevent in view of the fermionic gap and therefore disregarded throughout.

For excitations close to the K point we use a continuum Hamiltonian for fermions, too,

\begin{equation} \label{eq:HCont}
h_{\rm eff}(\v p) = c \v p \cdot \boldsymbol \sigma + m \sigma_z.
\end{equation}

Here, $m = w_0 + \sqrt{27} w_2, c = \sqrt{3} w_1/2$, and $\v p$ is the momentum relative to the K-point, $\boldsymbol \sigma = (\sigma_x, \sigma_y)$, where $\sigma_{x,y,z}$ are Pauli matrices in sublattice space. An analogous equation holds close to the K' point, but there the mass is $\bar m = w_0 - \sqrt{27} w_2$.

Using the effective continuum theory with the above assumption, the convoluted GF, Eq.~\eqref{eq:GF}, can be calculated exactly in the simplified case $c = v$ of identical speed for $e$-particles and Dirac fermions 
\begin{subequations}
\begin{align}
\mathbf G(i \epsilon, \v p )  &=  -{\frac{\sqrt{3}}{J}}\frac{\arctan\left (\frac{K}{r+m}\right)}{4\pi K} \notag \\
&\times \left [F_{r,m}(K)(i\epsilon + c \v p \cdot \boldsymbol \sigma) +2 m \sigma_z\right ] \label{eq:ExactGFa}
\end{align}
where $K = \sqrt{\epsilon^2 + c^2 p^2}$ and
\begin{align}
F_{r,m}(K) & = 1 - \frac{r - m}{r + m} \left [\frac{r +m}{K\arctan\left (\frac{K}{r +m} \right)} - \frac{(r + m)^2}{K^2} \right].
\end{align}
\label{eq:ExactGFtotal}
\end{subequations}

This result for a non-Fermi liquid GF is central to this work and has multiple implications. First, we remind the reader of the minimal energy associated to the creation of a physical fermion, which fractionalizes into $e$-particle (mass $r$) and orthogonal fermion (mass $m$). This is technically reflected in the branch cuts of the inverse trigonometric function entering the analytically continued $\mathbf G^R(E, \v p)$ and physically implies a continuum in the spectral weight for energies $E > r+m$, but a spectral gap below this energy, Fig.~\ref{fig:ContLimit} b).
Second, as can be readily guessed from the $r = m$ limit of the GF numerator of Eq.~\eqref{eq:ExactGFtotal}, there are zeros in the numerator of zeros with dispersion $\epsilon_\emptyset(\v p) = \pm \sqrt{m_\emptyset^2 + c^2 p^2}$.
Third, the case $r = m$ is very special, as the mass of zeros $m_\emptyset = r +m = 2r$ coincides with the minimal energy of the continuum of poles (the branch cut). For $r>m$ the minimal energy $m_\emptyset$ of GF zeros decreases and vanishes when $r/m \rightarrow \infty$, Fig.~\ref{fig:ContLimit} b).
However, for negative $r-m$, the cost to create an orthogonal fermion exceeds that of an $e$-particle, and the solutions to $\det [\mathbf G^R(E, 0)]=0$ become complex\cite{Suppmat}.
By invoking semiclassics, this also implies the absence of fermionic bound states or boundary zeros at smooth domain walls in $r$ describing ZTI-TI interfaces in real space. Fourth, to obtain the equivalent of Eq.~\eqref{eq:ExactGFtotal} for the K' point on needs to replace $m \rightarrow \bar m$ in front of $\sigma_z$ in the numerator and $m \rightarrow \vert \bar m \vert$ anywhere else. At $\bar m = 0$, this implies a direct OI-ZTI transition with a Dirac node of zeros and semiclassically a boundary zero at an OI-ZTI interface, Fig.~\ref{fig:ContLimit} d). Fifth, it is worthwhile to highlight a very different behavior in the spectral weight  $- \text{Im}[\mathbf G^R(E, 0)]$ for $r>m$ and $r<m$, respectively. In particular, when $r>m$, i.e. the $e$-particle gap exceeds the fermionic gap, the spectral weight displays a rapid suppression as the energy is lowered from outside the spectral gap towards $\vert E \vert = r + m$. In contrast, for $r<m$ the spectral gap at at the K-point features ``coherence'' peak structures, which can be interpreted as a precursor of the previously mentioned finite quasiparticle weight on the confining side of the phase transition (i.e. when $r^2$ becomes negative). These features are in principle visible in the color coding of Fig.~\ref{fig:ContLimit} b), but exemplified more clearly for two cross cuts in Fig.~\ref{fig:ContLimit} c). Finally, one may use Eq.~\eqref{eq:ExactGFa} to evaluate the topological GF invariant, Eq.~\eqref{eq:Volovik}. Summation over valleys yields $N_3 = 1$ ($N_3 = 0$) in the topological (non-topological) case $w_2/w_0 >1/\sqrt{27}$ ($w_2/w_0<1/\sqrt{27}$) corresponding to a ZTI (OI). Crucially, the contribution to $N_3$ entirely stems from the bands of zeros, while ``Hubbard'' bands of poles drop out of $N_3$ \cite{Suppmat,BlasonFabrizio2023, WagnerSangiovanni2023}. As neither poles nor zeros display a gap closing as $h/J$ is increased, it follows that $N_3  = 1$ throughout the ZTI phase. 

\PRLSection{Transport signatures.} The present integrable model readily allows to access quantized transport coefficients (here the Hall conductance).  
{Importantly, unlike the \textit{Hatsugai-Kohmoto} approach,  the Hamiltonian Eq.~\eqref{eq:Htot} is local in real space and thus straightforwardly coupled to external U(1) gauge fields. }
Lattice models of the type of Eq.~\eqref{eq:Htot} encode ``orthogonal'' metals, semimetals or (topological) insulators\cite{NandkishoreSenthil2012,ZhongLuo2012,ZhongLuo2013,HohenadlerAssaad2018}. A defining characteristic of the latter is that two-particle GF (e.g. polarization bubble or current-current correlator) are exactly the same as in the corresponding non-fractionalized phase, despite  the fact that the single particle propagator (i.e. the two-point function) resembles a Mott insulator. Another way to state this is that diagrams associated to electromagnetic response are evaluated using $G(i \epsilon, \v p)$ rather than $\mathbf G(i \epsilon, \v p)$. To understand why, for simplicity consider the hopping term along a single link $c^\dagger_{\v x} Z_{\v x, \v x'} c_{\v x'} +H.c.$. Peierls substitution and derivative with respect to an external U(1) gauge field leads to a current operator $j \sim i c^\dagger_{\v x} Z_{\v x, \v x'} c_{\v x'} +H.c.$. This current operator commutes with $Q_{\v x_0}$ for any $\v x_0$, and, by consequence, gapped $e$-particle propagators are not entering the current-current correlators, Fig.~\ref{fig:GreensFunction} a).
As a corollary, it follows that the Hall conductance is~\cite{ZhongLuo2013}
\begin{equation}
    \sigma_H = \Theta(\sqrt{27} w_2 -w_0), \label{eq:SigmaH}
\end{equation}
irrespective of the value of $h/J$. Throughout the letter, including Fig.~\ref{fig:Schematics} b), we express $\sigma_H$ in units of the von-Klitzing constant.

What is the relationship between $N_3[\mathbf G]$, Eq.~\eqref{eq:Volovik}, and $\sigma_H$? By applying the Widom-Streda formula\cite{Widom1982,Streda1982} 
it has been shown\cite{BlasonFabrizio2023,GavenskyGoldman2023,ZhaoPhillips2023, SettySi2023b} on general grounds that $\sigma_H[\mathbf G] = N_3[\mathbf G] + 2\pi \partial_B I_L[\mathbf G] \vert_{B = 0}/e$, where $I_L[\mathbf G]$ is the Luttinger integral\cite{Suppmat}. Note that in the context of this proof, $\sigma_H[\mathbf G]$ is evaluated with the full GF $\mathbf G$, even though, as argued above, for the present system the physical Hall response $\sigma_H$ is evaluated using the free fermion GF $G$. We explicitly evaluate the Widom-Streda formula at $J = 0$ to deduce $\sigma_H[\mathbf G] = \Theta(\sqrt{27} w_2 -w_0)$\cite{Suppmat}. Given that we independently calculated $N_3[\mathbf G] =0$ in this regime, one may conclude $2\pi \partial_B I_L[\mathbf G] \vert_{B = 0}/e =  \Theta(\sqrt{27} w_2 -w_0) $ and that $\sigma_H[\mathbf G] = \sigma_H$.

\PRLSection{Discussion and Outlook.} In summary, we have studied an exactly soluble, local model of a fractionalized topological insulator~\cite{ZhongLuo2013,Stern2016}. Analytically controlled perturbation about the integrable limit reveals topological bands of GF zeros, which are sharp deep in the deconfining phase, but acquire a ``lifetime'' near the Higgs condensation. 

As an outlook devoted to the experimental detection of (topological) GF zeros, we discuss momentum conserving tunneling which has previously served as a crucial tool in determining many-body Fermi and Luttinger liquid physics in 2D (``planar tunneling''\cite{MurphyWest1995,BritnellNovoselev2012}) and 1D \cite{BarakYacoby2010}, respectively. In a device with momentum conservation such as Fig.~\ref{fig:ContLimit} d), the decay rate $ \Gamma (\epsilon^{\rm lead}_{\v p}, \v p)$ of electrons in the upper lead determines the current across the junction. We use the notation $\epsilon^{\rm lead}_{\v p},G^R_{\rm lead}$ for dispersion and GF of lead electrons, respectively, consider identical top and bottom leads and, for simplicity, scalar Green's functions. As we argue in the following \cite{Suppmat}, the planar tunneling rate and, by consequence, the tunneling conductance $dI/dV$ display characteristic dips in the presence of GF zeros. 

To lowest order $ \Gamma (\epsilon^{\rm lead}_{\v p}, \v p) = - t^2 \text{Im} \mathbf G^R(\epsilon_{\v p}^{\rm lead}, \v p) $ which vanishes for energies below the spectral gap of the central layer of the device. In this case, the leading elastic contribution is $\Gamma (\epsilon_{\v p}, \v p) = - t^4 [\text{Re} \mathbf G^R(\epsilon_{\v p}^{\rm lead}, \v p)]^2  \text{Im} G^R_{\rm lead}(\epsilon^{\rm lead}_{\v p}, \v p)$, which however also vanishes when the lead electron energy and momentum $(\epsilon^{\rm lead}_{\v p}, \v p)$ coincide with energy and momentum of zeros in the central layer. 
For 2D planar tunneling, these kinematic constraints might admittedly be challenging to fine-tune experimentally. However, 1D momentum conserving tunneling in a sandwich (chiral edge state -- ZTI edge zero -- chiral edge state) is practical, since it can be expected that TI edge state and ZTI edge zero dispersions cross each other at a given energy $E_*$. While momentum-conserving co-tunneling in the absence of edge zeros implies a finite differential conductance at $\mathcal O(t^4)$, the same experiment shows a suppressed signal at the characteristic voltage $E_*/e$ in the presence of topological edge zeros. 

\PRLSection{Note added:} \textit{During the completion of this manuscript we became aware of a related work~\cite{WagnerSangiovanni2023b}; where there is overlap our results agree.}

\PRLSection{Acknowledgements:}
It is a pleasure to thank A.~Blason, J.~Cano, A.~Chubukov, L.~Classen, M.~Fabrizio, F.~Pollmann, M.~Vojta, S.~Sur for useful discussions on the topic. The authors are particularly indebted to G.~Sangiovanni, N.~Wagner for sharing their unpublished results prior to posting them on the arXiv. CS and EJK acknowledge hospitality by the Kavli Institute for Theoretical Physics, where this work was initiated. This research was supported in part by grants NSF PHY-1748958 and PHY-2309135 to the Kavli Institute for Theoretical Physics (KITP).
UFPS was supported by the Deutsche Forschungsgemeinschaft (DFG, German Research Foundation) through a Walter Benjamin fellowship, Project ID 449890867, and the DOE office of BES, through award number DE-SC0020305.

\bibliography{ZerosToricCode}

\clearpage

\setcounter{equation}{0}
\setcounter{figure}{0}
\setcounter{section}{0}
\setcounter{table}{0}
\setcounter{page}{1}
\makeatletter
\renewcommand{\theequation}{S\arabic{equation}}
\renewcommand{\thesection}{S\arabic{section}}
\renewcommand{\thefigure}{S\arabic{figure}}
\renewcommand{\thepage}{S\arabic{page}}

\begin{widetext}
\begin{center}
Supplementary materials on \\
\textbf{"Topological Green's function zeros in an exactly solved model and beyond"}\\
{Steffen Bollmann$^1$, 
{Chandan Setty $^2$,} 
{Urban~F.~P.~Seifert $^3$}, 
{Elio J. K\"onig}$^1$}\\
{%
$^1$ Max-Planck Institute for Solid State Research, 70569 Stuttgart, Germany
}\\%
{%
$^2$Department of Physics and Astronomy, Rice Center for Quantum Materials, Rice University, Houston, Texas 77005, USA
}\\%
{%
$^3$Kavli Institute for Theoretical Physics, University of California, Santa Barbara, CA 93106, USA
}%
\end{center}
\end{widetext}

These supplementary materials contain a summary of all definitions employed in this manuscript, Sec.~\ref{sec:SM:Defs}, details on the derivations of the Green's function in two different limits of the fractionalized phase, Sec.~\ref{sec:SM:GFFrac}, a discussion of the Widom-Streda formula, Sec. \ref{sec:SM:Widom}, of the Green's function based topological invariant Sec.~\ref{sec:SM:Volovik}, and about momentum conserving tunneling probes Sec.~\ref{sec:SM:MomCons}.

All citations refer to the bibliography of the main text.

\section{Definitions and Conventions}
\label{sec:SM:Defs}

\subsection{The local lattice limit} \label{App:FTtriangularLattice}

We consider the thermodynamic limit and treat the lattice momentum $\textbf k$ as a continuous variable. This means, in particular, the sum 
\begin{equation}
    \lim_{N\rightarrow \infty} \frac{1}{N} \sum_{\textbf k} \rightarrow \frac{1}{\text{Vol B.Z.}} \int_{\text{B.Z.}} \text{d}\textbf k
\end{equation}
becomes an integral. We chose the primitive lattice vectors
\begin{equation}
    \textbf n_1 = \frac{1}{2}
    \begin{pmatrix}
    1 \\ \sqrt{3}
    \end{pmatrix} \quad\text{and}\quad
    \textbf n_2 = \frac{1}{2}
    \begin{pmatrix}
        -1 \\ \sqrt{3}
    \end{pmatrix}
\end{equation}
to be normalized. Thus the volume of the first Brillouin zone evaluates to $\text{Vol B.Z.}=\frac{2}{\sqrt{3}}(2\pi)^2$. The $\textbf k$ integration becomes $\frac{\sqrt{3}}{2}\int (\text{d} k)$ where we use the shorthand notation $(\text{d}k)=\frac{\text{d}\textbf k}{(2\pi)^2}$. 

In the main text, we use the notation $\textbf x$ for vertices of the honeycomb lattice. Here we use $\textbf{x}=(\textbf r, s)$ where $\v r$ is a Bravais lattice vector and $s\in\{A, B\}$ describe the sublattice degrees of freedom. The Fourier transformation of the annihilation operator is defined as
\begin{equation}
    c_{\textbf r, A/B} = \sqrt{\frac{\sqrt{3}}{2}}\int (\text dk) \text e^{i \textbf k \cdot\textbf r}c_{A/B}(\textbf k).
\end{equation}
from which directly follows that 
\begin{equation}
    G_{s, s'}(\tau, \textbf r) = \frac{\sqrt{3}}{2}\int (\text d k) \text e^{i\textbf k \cdot \textbf r} G_{s, s'}(\tau, \textbf k),
\end{equation}
where $G_{s, s'}(\tau, \textbf r)=-\langle\mathcal T\left[ c^{\phantom{\dagger}}_{\textbf r, s}(\tau)c^\dagger_{0, s'}(\tau)\right]\rangle$ is the imaginary time ordered GF and $s, s'\in\{A, B\}$.

\subsection{The Model}

For further references, a more detailed representation of the model Hamiltonian \eqref{eq:Htot}, $H  =H_{\rm TC}+ H_{\rm f} + \delta H$ is worthwhile:

    \begin{subequations}
\begin{widetext}
\begin{align}
    H_{\rm TC} &= - K \sum_{\varhexagon} B_{\varhexagon} - h \sum_{\v r} [Q_{\v r}^A+Q_{\v r}^B],\\
    H_{\rm f} & = - w \sum_{\v r} \left [ \sum_{\v r' \in \{\v r, \v r- \hat n_1, \v r - \hat n_2 \}} c^\dagger_{\v r,A} Z_{\v b_{(\v r,A), (\v r',B)}} c_{\v r',B} + H.c.   \right ] + w_0 \sum_{\v r} \sum_s (-1)^s c^\dagger_{\v r, s}c_{\v r, s} \notag \\
    &- i w_2 \sum_{\v r} \left [\sum_{\substack{\v r' \in \\ \{ \v r + \hat n_1 - \hat n_2, \v r- \hat n_1. , \v r+ \hat n_2\} }} c^\dagger_{\v r,A} \prod _{\substack{ \v  b\in \\\gamma_{(\v r,A), (\v r',A) } }}Z_{\v b} c_{\v r', A}-\sum_{\substack{\v r' \in \\ \{ \v r + \hat n_1 - \hat n_2, \v r- \hat n_1. , \v r+ \hat n_2\} }} c^\dagger_{\v r,B} \prod _{\substack{\v  b\in \\ \gamma_{(\v r,B), (\v r',B) } }}Z_{\v b} c_{\v r', B}- H.c. \right ] \label{eq:Hfermions},\\
    \delta H & = - J \sum_{\v b} Z_{\v b},
\end{align}
\end{widetext}\end{subequations}
where
$
B_{\varhexagon} = \prod_{\v b \in \partial \varhexagon} Z_{\v b} ,
Q_{\v r}^s = (-1)^{\hat n_{\v r}^s}\prod_{\v b \in \partial (\v r,s)} X_{\v b}$

and $X_{\textbf b},Z_{\textbf b}$ are Pauli Matrices living on the bond $\textbf b$. The path $\gamma_{(\v r, s), (\v r', s')}$ connects two points on the shortest path.

Note that each unit cell contains three qubits and three integrals of motion (at $J =0$, that is), i.e. $B_{\varhexagon}, Q_{\v r}^A, Q_{\v r}^B$. 

\subsection{Ground state in the exactly solvable limit $J=0$}

As shown in \cite{KoenigTsvelik2020}, the ground state of a TC coupled to fermions can be constructed by projecting the tensor product of a fully polarized gauge field state with the fermionic sector.

To explicitly construct the Fermi sea, we need to switch to Fourier space. Starting from the gauge $Z_{\v b} = 1$ everywhere, we find
\begin{subequations}
\begin{align}
       H_{\rm f} = \int_{\rm BZ} (\text dk) \left (c^\dagger_A(\v k), c_B^\dagger(\v k) \right ) h(\v k) \left (\begin{array}{c}
         c_A(\v k)  \\
         c_B(\v k) 
    \end{array} \right)
\end{align}
Here, we used the Fourier transform as explained in detail in App.~\ref{App:FTtriangularLattice}. The single-particle Hamiltonian has the form
\begin{equation} \label{eq:h-fermion}
    h(\v k) = \left ( \begin{array}{cc}
        w_2 m(\v k) + w_0 & - w_1s(\v k) \\
        -w_1 s^*(\v k) & - w_2 m(\v k) - w_0
    \end{array}\right)
\end{equation}
with 
\begin{align}
    s(\v k) &= [e^{-i \v k \cdot \v n_1} + e^{-i \v k \cdot \v n_2} + 1]\\
    m(\v k) & = \left [ \sin(\v k \cdot \v n_{1-2})-\sin(\v k \cdot \v n_{1}) +\sin(\v k \cdot \v n_{2}) \right]
\end{align}
\end{subequations}
We will also need the energy 
\begin{equation}
    \epsilon(\v k)  = \sqrt{w_1^2|s(\v k)|^2 + (w_2 m(\v k)+w_0)^2}.
\end{equation}

\section{Fermionic Green's Function in the presence of topological order}
\label{sec:SM:GFFrac}

According to the discussion in the main text and a comprehensive discussion in Ref.~\cite{KoenigTsvelik2020}, the full GF for fermionic excitations in the OTI and ZTI is given as the component-wise product of the fermionic GF $G_{s, s'}(\tau, \v r, \v r')$ and the $e$-particle GF $D_{s, s'}(\tau, \v r, \v r')$, that is 
\begin{align}
    \textbf G_{s, s'}(\tau, \v r, \v r') = G_{s, s'}(\tau, \v r, \v r')D_{s, s'}(\tau, \v r, \v r') 
\end{align}
which translates to the convolution 
\begin{equation} \label{eq:convolution}
    \textbf G_{s, s'}(i\epsilon,\textbf k) = \frac{\sqrt{3}}{2}\int (\text dQ)  D_{s, s'}(i\omega, q)G_{s, s'}(i(\epsilon- \omega), \textbf k - \textbf q)
\end{equation}
in momentum/Matsubara frequency space, and $(dQ)  = d\omega d^2q/(2\pi)^3$.

\subsection{Propagator of orthogonal fermions}

As explained in the main text, the GF of the orthogonal fermions in momentum space is given as
\begin{equation}
    G(i\epsilon, \v k) = [i\epsilon - h(\v k)]^{-1} = - \frac{i\epsilon + h(\v k)}{\epsilon^2 + \epsilon(\v k)^2} .
\end{equation}

\subsection{Propagator of $e$-particles}

In the limit of vanishing $J$, the $e$-particle can not move over the lattice, and the GF becomes ultra-local and takes the simple shape 
\begin{equation}
    D^{(0)}_{s, s'}(\tau, \v r, \v r') = \text e^{-2h|\tau|}\delta_{s, s'}\delta_{\v r, \v r'}
\end{equation}
or 
\begin{equation}
    D^{(0)}_{s, s'}(i\omega, \v q) = \frac{4h}{\omega^2 + 4h^2}. 
\end{equation}

If non-Gaussian fluctuations of the gauge field are subleading (an estimate is given in Sec.~\ref{sec:est-gl}, below), the $e$-particle propagator can be approximated by the recursion formula 
\begin{align}
     D_{s_f,s_i} \left (\v r_f, i \omega \right) & = D^{(0)}(i \omega) \delta_{s_f, s_i} \delta_{\v r_f,\v r_i} \notag \\
    &+ J \sum_{\langle(\v r, s),(\v r_f, s_F)\rangle} D_{s,s_i}(\v r, i \omega)D^{(0)}(i\omega),
    \label{subEq:DRecursion}
\end{align}
where the brackets $\langle ... \rangle$ stand for nearest neighbors.
It incorporates all possible paths the particle can take from its initial position $(r_i, s_i)$ to the final position $(r_f, s_f)$ that do not intersect themselves. A more detailed discussion can be found in Ref.~\cite{KoenigTsvelik2020, RoyKoenig2023}. 

In momentum space, the recursion equation \eqref{subEq:DRecursion} takes the shape 
\begin{equation}
    D(i\omega, \v q) = D^{(0)}\left[\bm 1+ J \begin{pmatrix}
      0 &  s(\v q) \\ s^*(\v q) & 0
    \end{pmatrix} D(i\omega, \v q) \right]
\end{equation}
and we obtain a formal solution
\begin{equation} \label{eq:eprop}
\begin{split}
     \, & D(i\omega, \v q) = 4h \left [\omega^2 + 4 h^2 - 4 hJ \left (\begin{array}{cc}
        0   &  s(\v q)  \\
     s^*(\v q) & 0
    \end{array} \right)\right]^{-1} \\
    & = \frac{4h}{\omega^2 + 4h^2 - (4hJ)^2|s(\v q)|^2}\begin{pmatrix}
        \omega^2 + 4h^2 & 4hJs(\v q) \\ 4hJs^*(\v q) & (\omega^2 + 4h^2)
    \end{pmatrix}
    \end{split}
\end{equation}
for the $e$-particle propagator. The matrices act in sublattice space. 

In the following, we present the calculation of $\mathbf G$, first, in a perturbative expansion for $h/J \gg 1$ and, second, in a continuum limit when $J\sim h$ where we, however, stay far enough from the Higgs transition where non-Gaussian gauge field fluctuations become important. 

\subsection{Pertuabative GF for $h/J \gg 1$}

We now expand Eq.~\eqref{eq:eprop} up to the second order in $J$ and insert it order by order in the convolution \eqref{eq:convolution}. 

For $w_0=0$, the dispersion relation $\epsilon(\v k)$ would be $C_6$ symmetric, simplifying further calculations. Any finite $w_0$ reduces the symmetry to $C_3$. However, if we stay within the hierarchy of limits $h\gg w_1 \gg w_2 \gg w_0$ (in the top left of the phase diagram \ref{fig:Schematics}b) and $h \gg \epsilon$ we can expand the dispersion relation
\begin{equation}
    \epsilon(\v k) = \sqrt{\underbrace{w_1^2|s(\v k)|^2 + w_2^2 m(\v k)^2 + w_0^2}_{\xi^2(\v k)} + 2w_0w_2 m(\v k)}.
\end{equation}
Note that the term $2w_0w_2 m(k)$ is, for any $\v k$, small compared to $\xi^2(\v k)$. Thus, we can expand $\epsilon(\v k)$ in powers of $w_0w_2 m(\v k)$ and obtain up to first order 
\begin{equation}
    \epsilon(\v k) \simeq  \xi(\v k) + \frac{w_0 w_2 m(\v k)}{\xi(\v k)}.
\end{equation}
We split $\epsilon(\v k)$ into a dominating $C_6$ symmetric and a subleading $C_6$ anti-symmetric part. 

It will be useful to introduce the function 
\begin{equation}
    g(x, y) = \frac{1}{x^2 + y^2}.
\end{equation}

\subsubsection{GF to order $J^0$}

The integral
\begin{equation}
    \begin{split}
    \mathbf G(i\epsilon)|_{J^0} & = -\frac{\sqrt{3}}{2}\int(\text d q)(\text d \omega)D^{(0)}(i\omega) \\ & \times\frac{i(\epsilon-\omega) + (w_2 m(\v k - \v q)+w_0)\sigma_z}{(\epsilon -\omega)^2 + \epsilon^2(\v k - \v q)}
        \end{split}
\end{equation}
gives the zeroth order contribution to the GF. Performing the $\omega$ integration and applying the shift $\v q \rightarrow - \v q + \v k$ yields 
\begin{equation}
    \begin{split}
    \mathbf G(i\epsilon)|_{J^0} & = -\frac{\sqrt{3}}{2}\int(\text d q) \Bigg(i \epsilon g(\epsilon, \epsilon(\v q) + 2h) \\ & + \left. \left( 1+\frac{2h}{\epsilon(\v q)}\right)g(\epsilon, \epsilon(\v q)+2h)(w_2 m(\v q) + w_0)\sigma_z \right).
    \end{split}
\end{equation}
Now, we use that (in the assumed limit) $h$ is the dominant energy scale, and approximate the function $g(\epsilon, \epsilon(\v k))\rightarrow\frac{1}{4h^2}$. This approximation will be used throughout the rest of this section.
Thus, the first part simply contributes  $\frac{-i\epsilon}{4h^2}$. In the second part, we only keep the term in leading order in $1/h$. 

Also, we expand the function $g(\epsilon, \epsilon(\v q)+2h)/\epsilon(\v q)$ in the subleading $C_6$ anti-symmetric part, according to the procedure outlined in the last section. We obtain 
\begin{equation}
    \begin{split}
    \frac{g(\epsilon, \epsilon(\v q + 2h))}{\epsilon(\v q)} & \approx \frac{g(\epsilon, \xi(\v q) + 2h)}{\xi(\v q)} \\ & - \frac{w_0 w_2 g(\epsilon, \xi(\v q))}{\xi^3(\v q)}m(\v q) \\ & - 2\frac{w_0 w_2 g^2(\epsilon, \xi(\v q)+ 2h)}{\xi^{2}(\v q)}\left(1 + \frac{2h}{\xi(\v q)}\right) m(\v q).
    \end{split}
\end{equation}
All anti-symmetric terms (i.e., those are proportional to $m(\v q)$) are either of higher order in $1/h$ or are suppressed by additional factors $w_0/w_1$ or $w_2/w_1$ compared to the symmetric part. Therefore, we also ignore these terms. Applying the approximations yields
\begin{equation}
    \begin{split}
        \mathbf{G}|_{J^0}(i\epsilon) & = - \frac{i\epsilon}{4h^2} - \frac{\sqrt{3}}{2}\int (\text d q) \frac{2hg(\epsilon, \xi(\v q)+2h)}{\xi(\v q)} w_0 \sigma_z \\ & \approx - \frac{i\epsilon}{4h^2} - \frac{1}{hw_1} \underbrace{ \sqrt{3}\int (\text d q) \frac{1}{|s(\v q)|}}_{\mathcal{C}_0}\sigma_z \\ & = -\frac{1}{4h^2}(i\epsilon + \frac{h}{w_1}\mathcal{C}_0\, w_0\sigma_z),
    \end{split}
\end{equation}
where $\mathcal C_0 \approx 1.8$. Here, we first used that, due to symmetry, 
\begin{equation}
    \int (\text d q)g(\epsilon, \xi(\v q)+2h)m(\v q) = 0
\end{equation}
and, second, that in the limit $w_1\gg w_2 \gg w_0$ the scaling of $\xi(\v q)$ is essentially given by $w_1$. Therefore, we can approximate $\xi(\v q)\approx w_1 |s(\v q)|$.

\subsubsection{GF to order $J^1$}

The integral
\begin{equation}
    \begin{split}
    \mathbf G(i\epsilon, \v k)|_{J^1} & = -J\frac{\sqrt{3}}{2}\int(\text d q)(\text d \omega)(D^{(0)}(i\omega))^2\\ & \times \begin{pmatrix}
        0 & \frac{-w_1s(\v q)s(\v k-\v q)}{(\epsilon - \omega)^2 + \epsilon^2(\v k - \v q)} \\ \frac{s^*(\v q)s^*(\v k-\v q)}{{(\epsilon - \omega)^2 + \epsilon^2(\v k - \v q)}} & 0
    \end{pmatrix}
        \end{split}
\end{equation}
gives the first order contribution to the GF. Using the identity 
\begin{equation}
    (D^{(0)}(i\omega))^2 = - \frac{h}{2}\partial_h\frac{1}{h} D^{(0)}(i\omega)
\end{equation}
the frequency integration is analogous to the one for the zeroth order. Performing the integration and only keeping terms that are leading in $1/h$ we obtain 
\begin{equation}
    \begin{split}
    \mathbf G(i\epsilon, \v k)|_{J^1} & =  J h\frac{ \sqrt{3}}{2}\partial_h \int (\text d q) \frac{g(\epsilon, \xi(\v q)+2h)}{\xi(\v q)} \\ & \times
    \begin{pmatrix}
        0 & -w_1s(\v q)s(\v k-\v q) \\ -w_1s^*(\v q)s^*(\v k-\v q) & 0
    \end{pmatrix}\\
    & = J h \frac{\sqrt{3}}{2}\begin{pmatrix}
        0 & -w_1 s(\v k) \\ -w_1 s^*(\v k)
    \end{pmatrix}
    \\ & \times \partial_h\int(\text d q) \frac{g(\epsilon, \xi( \v q)+2h)}{\xi(\v q)}(1+2\cos(\v q\cdot \v n_{1/2})) \\ &
    \approx -\frac{1}{4h^2}J \mathcal C_0\mathcal{C}_1\begin{pmatrix}
        0 & -s(\v k) \\ -s^*(\v k),
    \end{pmatrix}
    \end{split}\label{eq:g-first-order}
\end{equation}
where 
\begin{equation}
   \mathcal C_0 \mathcal{C}_1 \equiv \frac{\sqrt{3}}{4}\int(\text d q)\frac{1+2\cos(\v q \cdot \v n_{1/2})}{|s(\v q)|}\approx 0.26.
\end{equation}
To achieve the second equality in Eq.~\eqref{eq:g-first-order}, we expand the product $s(\v q)s(\v k - \v q)$ and use the $C_6$ symmetry of the integrand. $\v n_{1/2}$ denotes that both $\v n_1$ and $\v n_2$ can be used since they yield the same numerical value  (again due to the $C_6$ symmetry).

\subsubsection{GF to order $J^2$}

The integral 
\begin{equation}
    \begin{split}
       \mathbf{G}(i\epsilon, \v k)|_{J^2} & =  - \frac{\sqrt{3}}{2}J^2\int(\text d q)(\text d \omega) (D^{(0)}(i\omega))^3\\ & \times |s(q)|^2\frac{w_2 m(\v k - \v q)+w_0}{(\epsilon -\omega)^2+\epsilon(\v k - \v q)}
    \end{split}
\end{equation}
gives the second-order contribution to the GF. Using a similar trick as for the first-order, that is
\begin{equation}
    (D^0(i\omega))^3 = \underbrace{\frac{1}{8}\left(h\partial_h^2 - \partial_h\right)}_{\mathcal D_h}\frac{1}{h}D^{(0)}(i\omega)
\end{equation}
enables a straightforward frequency integration. We obtain 
\begin{equation}
    \begin{split}
        \mathbf{G}(i\epsilon, \v k)|_{J^2} & = -\frac{\sqrt{3}}{2}\mathcal{D}_h \frac{1}{h} \int (\text d q) |s(\v k - \v q)|^2 m(\v q)\\  & \times\frac{g(\epsilon, \xi(\v q)+2h)}{\xi(\v q)}\sigma_z
        \\ &  \approx -\frac{\sqrt{3}}{{2}}\mathcal D_h \frac{m(\v k)}{2h^2} \int(\text d q) \notag \\ &\times \frac{2\sin^2(\v q\cdot\v n_{1/2})\sin^2(\frac{\v q\cdot \v n_{2/1}}{2})}{w_1|s(\v q)|}\sigma_z \\ & = -\frac{1}{4h^2} \frac{J^2}{hw_1} \mathcal C_0 \mathcal{C}_2 m(\v k)\sigma_z,
    \end{split}
\end{equation}
where
\begin{equation}
    \mathcal C_0 \mathcal{C}_2 = 4\sqrt{3}\int(\text d q)\frac{\sin^2(\v q\cdot\v n_{1/2})}{|s(\v q)|}\sin^2\left(\frac{\v q\cdot \v n_{2/1}}{2}\right)\approx 2.4.
\end{equation}
Here, we used trigonometric identities again to expand $|s(\v k-\v q)|^2m(\v q)$ and the $C_6$ symmetry to simplify the expression under the integral sign. 

\subsubsection{Full Result}

Adding all calculated orders yields the final expression for the GF:
\begin{subequations}
\begin{equation}
\begin{split}
     \mathbf{G}(i\epsilon, \v k) & =  \mathbf{G}(i\epsilon, \v k)|_{J^0}  + \mathbf{G}(i\epsilon, \v k)|_{J^1} + \mathbf{G}(i\epsilon, \v k)|_{J^2} \\
    & = \frac{-1}{4h}\left(i\epsilon + \mathcal C_0\frac{h}{w_1} h_{\emptyset}\right),
\end{split}
\end{equation}
where 
\begin{equation}
    h_{\emptyset} = \begin{pmatrix}
    w_0 + \frac{J^2}{h^2} w_2\mathcal C_2  m(\v k) & -\frac{J}{h}w_1\mathcal C_1 s(\v k) \\ - \frac{J}{h} w_1\mathcal C_1 s^*(\v k) & -w_0 - \frac{J^2}{h^2} \mathcal C_2 w_2 m(\v k)
    \end{pmatrix} .
\end{equation}
\end{subequations}
This is the origin of equation \eqref{eq:HZero}.

\subsection{Estimate of Ginzburg-Levanyuk number} \label{sec:est-gl}

For the next section, in which we employ a continuum approximation of both e-particle and fermionic propagators it is useful to discuss the limits of applicability of the present theories without non-Gaussian fluctuaions.

The confinement transition is described by an effective $\phi^4$ theory 
\begin{equation}
\mathcal L  = \phi[-\partial_\tau^2 - \nabla^2+r^2]\phi + \lambda \phi^4 
\end{equation}
Note that $[r] = E,[\phi] = E^{-1+D/2}$, $[\lambda] = E^{4-D}$
One may estimate the regime of negligible non-Gaussian fluctuations as 
\begin{equation}
r>\lambda^{- \frac{1}{D-4}}.
\end{equation}
Our bare estimates of coupling constants are\cite{KoenigTsvelik2020}  
are $r = \sqrt{h^2-3hJ}$, $\lambda = J^3/h^2$, leading to
\begin{equation}
\sqrt{h^2 - 3hJ} > J^3/h^2.
\end{equation}

\subsection{$e$-particle and orthogonal fermion GFs in continuum limit}

In the case where $J\sim h$ we can find an exact expression of $\textbf G_{s, s'}(i\epsilon, \v k)$ for the low energy/momentum limit. 
First, we diagonalize the $e$-particle propagator \eqref{eq:eprop} and obtain 
\begin{equation}
\begin{split}
    D(i\omega, \v q) & = \frac{4h}{\omega^2 + 4h^2 - 4hJ|s(q)|}\begin{pmatrix}
        1 & 0 \\ 0 & 0
    \end{pmatrix} 
    \\ & + \frac{4h}{\omega^2 + 4h^2 + 4hJ|s(q)|}\begin{pmatrix}
        0 & 0 \\ 0 & 1 
    \end{pmatrix} \\
     & \approx \frac{4h}{\omega^2 + 4h^2 - 4hJ|s(q)|}\begin{pmatrix}
        1 & 0 \\ 0 & 0
    \end{pmatrix},
\end{split}
\end{equation}
where we only keep the lower e-particle band in the second line. The transformation matrix reads
\begin{equation}
    T = \frac{1}{\sqrt{2}}
    \begin{pmatrix}
        \sqrt{\frac{s(q)}{s^*(q)}} & -\sqrt{\frac{s(q)}{s^*(q)}} \\ 1 & 1
    \end{pmatrix}
    \approx \frac{1}{\sqrt{2}}\begin{pmatrix}
        1 & -1 \\ 1 & 1
    \end{pmatrix}.
\end{equation}
As we will be interested at the lowest $e-$particle excitaitons (near the $\Gamma$ point), we expanded the transformation matrix near $\v q=0$. Transforming the propagator for the lower band back into the sublattice basis using $T$ yields 
\begin{equation}
    D(i\omega, \v q) \simeq \frac{4h}{\omega^2 + 4h^2 - 4h J |s(\v q)|}\begin{pmatrix}
        1 & 1\\ 1 & 1
    \end{pmatrix}.
\end{equation}
Since we multiply the $e$-particle and fermion GF component-wise, the matrix structure of $D(i\omega, \v q), $ allows us to treat it as a scalar in the subsequent calculation.  

In the long wavelength limit, we can approximate $\vert s(\v q)\vert$ as a parabola (see Fig.~\ref{fig:ContLimit} of the main text) and obtain for the $e$-particle propagator
\begin{equation}
    D(i\omega, \v q) = \frac{4h}{\omega^2 + r^2 + v^2 q^2},
\end{equation}
where $v=\sqrt{hJ}$, $r^2 = 4h(h-3J)$ and $q=|\v q|$.

The GF for the fermions in the long wavelength limit can be calculated straightforwardly using the continuum limit of \eqref{eq:h-fermion} and reads
\begin{equation}
    G(i\epsilon, \v p) = (i\epsilon - h_\text{eff}(\v p))^{-1} = -\frac{i\epsilon + h_\text{eff}(\v p)}{\epsilon^2 + m^2 + c^2 p^2},
\end{equation}
where $m = w_0+3\sqrt{3}w_2$, $c=\frac{\sqrt{3}}{2}w_1$,$p = |\v p|$ and $\v p$ is the momentum relative to the K point.

\subsection{Full GF in continuum's limit}

The full GF in the continuum limit is best calculated in real space. It is straightforward to show that the $e$-particle propagator in real space reads (we here use the notation $\v x$ for the continuum coordinate to avoid confusion with the e-particle gap $r$)
\begin{equation}
    D(\tau, \v x) = \frac{4h}{v^2}Y_r(X_v),
\end{equation}
where 
\begin{equation}
    Y_r(X_v) = \frac{\text e^{-r X_v}}{4\pi X_v} = \int (\text{d}\omega)(\text{d}q) \frac{v^2 \text e^{i\v q \cdot \v x}\text e^{-i\omega \tau}}{\omega^2 + r^2 + v^2 q^2 }
\end{equation}
and $X_v = \sqrt{\tau^2 + |\textbf r|^2/v^2}$.

The fermion GF can be similarly calculated:

\begin{equation}
\begin{split}
G(\tau, \v x) & = -\int (dp)(d\epsilon) e^{i \v p \cdot \v x}e^{-i \epsilon \tau} \frac{i \epsilon + h(\v p)}{\epsilon^2 + c^2 p^2 + m^2} \\
& = -\int (dp)(d\epsilon) e^{i (\v  p \cdot \v x + \epsilon \tau)} \frac{-i\epsilon + \v p\cdot\boldsymbol{\sigma}+m\sigma_z}{\epsilon^2 + m^2 + c^2 p^2} \\
& =  \frac{[\partial_\tau + i c \nabla \cdot \sigma - m \sigma_z]}{c^2}\int (dp)(d\epsilon) \frac{ c^2 e^{i (\v  p \cdot \v x + \epsilon \tau)}}{\epsilon^2 + m^2 + c^2 p^2}
\\ & = \frac{[\partial_\tau + i c \nabla \cdot \sigma - m \sigma_z]}{c^2} Y_m(X_c).
\end{split}
\end{equation}.

If restrict too the special case $v=c=1$ (i.e. $X_v=X_c=X=\sqrt{\tau^2 + |\v x|^2}$) the full GF in real space reads 
\begin{equation}
    \begin{split}
    \textbf G(\tau, \v x) & = 2\sqrt{3}h Y_r(X) \left[\partial_\tau + i\nabla\bm\sigma-m\sigma_z \right]Y_m(X) \\
    & = \sqrt{3}h \left[ \mathcal{D}_{\tau, \v x} + (r-m)\frac{\tau + i \v x \cdot\boldsymbol{\sigma}}{X} - 2m \sigma_z\right]\Upsilon_{r+m}(X)
    \end{split}
    \label{eqSub:RealSpaceGreensFunction}
\end{equation}
where $\Upsilon_{r+m}(X) = \frac{\text e^{-(r+m)X}}{(4\pi X)^2}$ and $\mathcal{D}_{\tau, \v x}=\partial_\tau + i\nabla\cdot\bm\sigma$.

Eq.~\eqref{eqSub:RealSpaceGreensFunction} can be Fourier transformed back into the momentum/frequency space. First, we calculate
\begin{equation}
    \begin{split}
    \Upsilon_{r+m}(K) & = \int \text{d}\v x \text d \tau \Upsilon_{r+m}(X) \text e^{-i\v x\cdot \v p}\text e^{i\epsilon\tau} \\ &= \frac{1}{4\pi}\int_0^\infty \text d X \text e^{-(r+m)X}\frac{\sin(KX)}{KX} \\ & = \frac{\arctan(K/(r+m))}{4\pi K}
    \end{split},
    \label{eqSub:FourieTransformUpsilon}
\end{equation}
where $K=\sqrt{\epsilon^2 + p^2}$.
We can use this result to calculate further 
\begin{equation}
    \begin{split}
     \, & \int \text d \v x\text d\tau \frac{(\tau + i\v x\cdot \bm \sigma)\text e^{-i\v x\cdot \v p}\text e^{i\epsilon\tau}}{X} \Upsilon_{r+m}(X) \\ & = -\underbrace{(i\partial_\epsilon + \nabla_{\v p}\cdot \bm \sigma)}_{\mathcal{D}_{\epsilon, \v p}} \int \text d \v x\text d\tau \frac{\text e^{-i\v x\cdot \v p}\text e^{i\epsilon\tau}}{X} \Upsilon_{r+m}(X) \\
     & = -\frac{1}{4\pi} \int_0^\infty \text d X \mathcal{D}_{\epsilon, \v p} \left(\frac{\sin(KX)}{KX^2}\right)e^{-(r+m)X}  \\ & = -\frac{\mathcal{D}_{\epsilon, \v p} K}{4\pi} \int_0^\infty \text d X \frac{1}{X} \partial_K\left(\frac{\sin(KX)}{KX}\right)e^{-(r+m)X} 
     \\ & = -\frac{i\epsilon + \v p\cdot\bm\sigma}{4\pi K^2} \int_0^\infty \text d X  \partial_X\left(\frac{\sin(KX)}{KX}\right)e^{-(r+m)X} \\ & = -\frac{i\epsilon + \v p\cdot\bm\sigma}{4\pi K^2}\left(\frac{r+m}{K}\arctan(K/(r+m))-1\right) \\ & = \frac{i\epsilon + \v p\cdot\sigma}{r+m}f(K/(r+m))\Upsilon_{r+m}(K), 
     \end{split}
     \label{eqSub:GFourieTrafoLongPart}
\end{equation}
where $f=(\frac{1}{x\arctan(x)}-\frac{1}{x^2})$. In the 4th equality we used that $\frac{1}{X}\partial_K \left(\frac{\sin(KX)}{KX}\right)=\frac{1}{K}\partial_X \left(\frac{\sin(KX)}{KX}\right)$ and $\mathcal{D}_{\epsilon, \v k} K = \frac{i\epsilon - \v k\cdot\bm\sigma}{K}$.

Combining the equations \eqref{eqSub:RealSpaceGreensFunction}, \eqref{eqSub:FourieTransformUpsilon} and \eqref{eqSub:GFourieTrafoLongPart} yields the exact GF in frequency space:
\begin{subequations}
\begin{align}
\mathbf G(i \epsilon, \v p )  &=  -{\frac{\sqrt{3}}{J}}\frac{\arctan\left (\frac{K}{r+m}\right)}{4\pi K} \notag \\
&\times \left [F_{r,m}(K)(i\epsilon + \v p \cdot \boldsymbol \sigma) +2 m \sigma_z\right]
\label{eqSub:FullGFContinuum}
\end{align}
where
\begin{align}
F_{r,m}(K) & = 1 - \frac{r - m}{r + m} \left [\frac{r +m}{K\arctan\left (\frac{K}{r +m} \right)} - \frac{(r + m)^2}{K^2} \right].
\end{align}
\end{subequations}
Note that we replaced $h$  by $1/J$ since we set $v=\sqrt{hJ}=1$. This is the origin of Eq.~\eqref{eq:ExactGFtotal}

\subsection{Spectral properties of GF in continuum limit}

In this section, we analytically continue the Green's function \eqref{eq:ExactGFtotal} to obtain the Green's function zeros from the retarded GF. We start analytical continuing $K$, which is a well-behaved function of $i\epsilon$ 
\begin{equation}
    \lim_{i\epsilon\rightarrow E + i0^+} K = i\underbrace{\sqrt{E^2 - p^2}}_{\mathcal K},
\end{equation}
where we assumed that $E>p$. Thus, the retarded Green's function reads
\begin{equation}
    \begin{split}
    \mathbf{G}^R(E, \v p) & = -\frac{\sqrt{3}}{J} \frac{\arctanh(\frac{\mathcal K}{r+m})}{4\pi \mathcal K} \\ & \times \left[F^R_{r,m}(E + \v p \cdot \bm \sigma) + 2m\sigma_z\right],
    \end{split}
\end{equation}
where 
\begin{equation}
    F^R_{r,m} = 1 - \frac{r-m}{r+m} \left[\frac{(r+m)^2}{\mathcal K^2} - \frac{r+m}{\mathcal K \arctanh(\frac{\mathcal K}{r+m})}\right].
\end{equation}

\subsubsection{Determinant of the GF}

The spectrum of the zero modes can be found by taking the determinant of the retarded GF and solving for 0, that is
\begin{equation}
    \begin{split}
    \det\mathbf G^R(\mathcal K) & = \frac{3}{J^2}\frac{\arctanh^2[\mathcal K/(r+m)]}{(4\pi \mathcal K)^2} \\ & \times \left[(F^R_{r, m})^2(\mathcal K)\mathcal K^2 - 4 m^2\right] = 0.
    \end{split}
\end{equation}
This equation can be recasted in the condition \begin{equation}
    \mathcal K = \frac{2m}{F_{r,m}(\mathcal K)}.
    \label{eqSub:ConditionGFzeros}
\end{equation}
We identify the solutions $\mathcal{K}_0$ to this equation with the restmass of zeros $\mathcal K_0 \equiv m_{\emptyset}$, since the spectrum of zeros near K, K' is given by
\begin{equation}
    E = \sqrt{p^2 + m_{\emptyset}^2}.
\end{equation}

\begin{figure}[t]
\includegraphics[width = .5\textwidth]{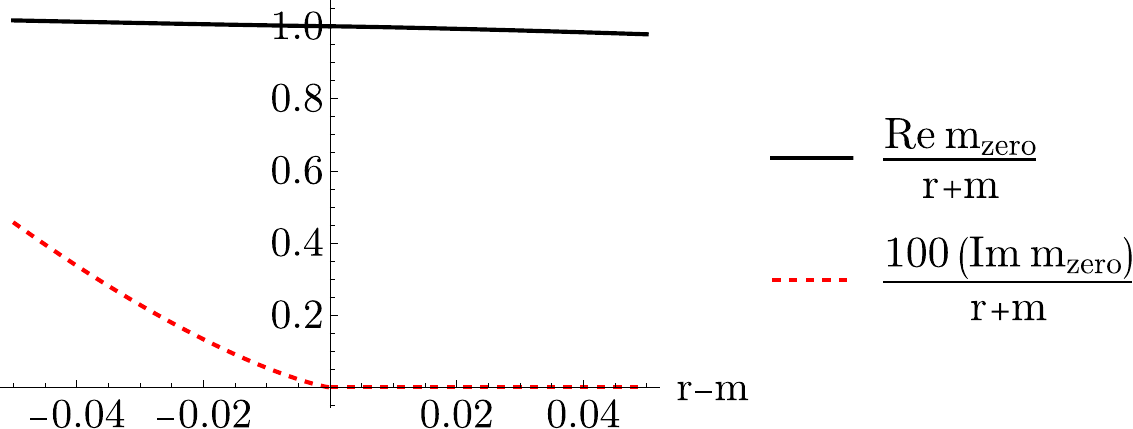}
\caption{Real and complex part of the zero's energy mass gap as a function of $r - m$. }
\label{fig:WhereZeros}
 \end{figure}

\subsubsection{Approximate solution near $r = m$}

We iteratively solve the equation  $\det \mathbf G^R(\mathcal K) = 0$ near $r = m$ and $\mathcal K = 1$. 
Using units in which $r + m = 1$ and $r- m \equiv \Delta$, the defining equation becomes to leading order in $\Delta$
\begin{equation}
\mathcal K = 1 + \frac{2 \Delta }{\ln \left(\frac{1- \mathcal K}{2}\right)} = 0.
\end{equation}
We see that the term with a large log in the denominator is subleading, so we solve the equation iteratively and we obtain
\begin{equation}
\mathcal K = 1+2 \frac{\Delta }{\log \left(-\frac{\Delta }{\log \left(-\frac{\Delta }{\log \left(-\frac{\Delta }{\log \left( \dots\right)}\right)}\right)}\right)}. \label{eq:SolNear1}
\end{equation}

The real and imaginary part of the GF zero mass as a function of $r-m$ at the K point is plotted in Fig.~\ref{fig:WhereZeros}. 
We can clearly see that the imaginary part of the mass suddenly becomes finite once the mass of the orthogonal fermions exceeds the e-particle mass, that is, $m>r$.

\section{Widom-Streda Formula}
\label{sec:SM:Widom}

In this section we demonstrate an independent calculation of $\sigma_H$ (in units of the von-Klitzing constant) using the Widom Streda formula

\begin{equation} \label{SM:eq:WidomStreda}
   \sigma_H  =  \left .\frac{2\pi}{ e } \frac{\partial n}{\partial B} \right \vert_{B = 0}.
\end{equation}
where the density is calculated with the full propagator $\mathbf G$. To be concrete, in the presence of a magnetic field it is convenient to express the density using the full real space GF
\begin{equation}\label{eq:Density}
    n[\mathbf G] = \frac{1}{\Omega} \int (\text d\epsilon) e^{i \epsilon 0^+} \sum_{\v r} \tr [\mathbf G(i \epsilon; \v r, \v r)],
\end{equation}
where $\Omega$ is the area of the system (total number of lattice sites). It is emphasized in the main text that the physical density $n$ is evaluated using $G$ in the present case, and the identification of $n[\mathbf G]$ with the density is a priori not guaranteed.

Using this expression, it was shown in Refs.~\cite{BlasonFabrizio2023,GavenskyGoldman2023} that 
\begin{equation}\label{eq:N3PlusLutt}
     \sigma_H[\mathbf G]  = N_3[\mathbf G] + \frac{2\pi}{e} \left .\frac{\partial I_L[\mathbf G]}{\partial B} \right \vert_{B = 0},
\end{equation}
where the Luttinger integral is 
\begin{equation}
    I_L [\mathbf G] = \frac{1}{\Omega}\int (\text d \epsilon) \text{Tr} [\mathbf G(i \epsilon) \frac{\partial \Sigma(i \epsilon)}{ i \partial  \epsilon}].
\end{equation}
Here, $\text{Tr}$ includes the trace over internal degrees of freedom as well as space/momentum space.

In the present model, a direct evaluation of $I_L$ is not natural, given that $\mathbf G$ is defined as a convolution and $\Sigma = G^{-1} - \mathbf G^{-1}$ is never explicitly evaluated. Instead, we here directly calculate $\sigma_H[\mathbf G]$ using the Widom Streda formula. Given that this result is independent from the previous section we thereby implicitly demonstrate that Eq.~\eqref{eq:N3PlusLutt} holds for the case of our system if $\sigma_H[\mathbf G]$ and $\sigma_H$ are identified.

We use for arbitrary eigenbasis $[\hat H \psi_n](\v r)  = E_n \psi_n(\v r)$ to evaluate Eq.~\eqref{eq:Density} at $J = 0$
\begin{align}
   \mathbf G(i \epsilon; \v r, \v r') & =  \sum_n \int (\text d \omega) D(i \omega) \sum_{n} \frac{\Psi_n(\v r) \Psi_n(\v r')}{i \epsilon - E_n} \notag \\
   & = \sum_n \frac{\Psi_n(\v r) \Psi_n(\v r')}{i \epsilon - \text{sign}(E_n) [\vert E_n \vert + 2 h]},
\end{align}
so that 
\begin{align}
     n[\mathbf G] &= \frac{1}{\Omega} \int (\text d\epsilon) e^{i \epsilon 0^+} \sum_n \frac{1}{i \epsilon - \text{sign}(E_n) [\vert E_n \vert + 2 h]}\\
     & = \frac{1}{\Omega} \sum_{n} \Theta (-E_n).
\end{align}
Note that $n[\mathbf G]$ at $h/J = \infty$ and $n$ are thus equal.

Finally, for the evaluation of $\sigma_H$ we calculate the difference of $\sigma_H$ at large and small $w_2/w_0$ and exploit that for $w_2 = 0$, $\sigma_H[\mathbf G] = 0$ by symmetry. We may readily use the continuum approximation Eq.~\eqref{eq:HCont} using canonical momenta $\Pi_\mu = p_\mu + e A_\mu$ and $m = w_0 + \sqrt{27} w_2$, $\bar m = w_0 - \sqrt{27} w_2$ We further use for $e>0, B>0$ that 
\begin{align}
b = \frac{1}{\sqrt{2  eB }} i [\Pi_x - i \Pi_y],
\end{align}
where $b, b^\dagger$ are regular bosonic ladder operators. Thus, using the notation $\ell_B = 1/ \sqrt{ eB  }$, we find
\begin{align}
h_K & = \left (  \begin{array}{cc}
m & -i \sqrt{2}\frac{c}{\ell_B} b \\ 
-i \sqrt{2} \frac{c}{\ell_B} b^\dagger & - m
\end{array} \right),\\
h_{K'} & = \left (  \begin{array}{cc}
\bar m & i \sqrt{2} \frac{c}{\ell_B} b^\dagger\\ 
-i \sqrt{2} \frac{c}{\ell_B} b & - \bar m
\end{array} \right).
\end{align}
The spectrum is ($\Omega_c^2 = \sqrt{2}{c}/{\ell_B}$)
\begin{align}
    E_n^{(K)} & = \begin{cases}
        -m, & n = 0, \\
        \pm \sqrt{m^2 + \Omega_c^2 n^2}, & n \neq 0,
    \end{cases}\\
     E_n^{(K')} & = \begin{cases}
        \bar m, & n = 0, \\
        \pm \sqrt{\bar m^2 + \Omega_c^2 n^2}, & n \neq 0.
    \end{cases}
\end{align}
Note that each Landau level has a degeneracy of $\vert eB \vert \Omega/2\pi$. At the topological transition $\bar m$ changes sign and the zeroth Landau level of the K' point plunges under the Fermi level leading to an increase in density
\begin{equation}
    n_{\bar m<0}[\mathbf G]-n_{\bar m>0}[\mathbf G] = \frac{ eB }{2\pi},
\end{equation}
i.e. $\sigma_{H}[\mathbf G] = \Theta(-\bar m)$.

\section{Green's function based topological formula}
\label{sec:SM:Volovik}

In this section we consider the contribution of a given valley to the quantized topological invariant, Eq.~\eqref{eq:Volovik} of the main text,
\begin{subequations}
\begin{align}
N_3 &= \frac{\pi}{3} \int (\text dK) q[\mathbf G(\vec K)],\\
q[\mathbf G(\vec K)] & = \epsilon^{\mu \nu \rho} \tr \{[\mathbf G^{-1}(\vec K)\partial_{K_\mu} \mathbf G(\vec K)] \notag \\ & \times [\mathbf G^{-1}(\vec K)\partial_{K_\nu} \mathbf G(\vec K)]\notag \\ &\times [\mathbf G^{-1}(\vec K)\partial_{K_\rho} \mathbf G(\vec K)]\}.
\end{align}
\label{eq:VolovikSM}
\end{subequations}
Here $(\text dK) = \text d^2k d\epsilon/(2\pi)^3$ and $\vec K = (\epsilon,k_x,k_y)$, and $\mu, \nu, \rho \in \{ 0,1,2\}$, and the totally antisymmetric Levi-Civita tensor $\epsilon^{\mu \nu \rho}$ is uniquely defined by setting $\epsilon^{012} = 1$.

Given  GF \eqref{eq:ExactGFtotal}, we rewrite the GF in the form
\begin{subequations}
\begin{align}
\mathbf G(\vec K)  &= \mathcal G(K) \tilde G(i \epsilon, \v k)\\ 
\tilde G(i \epsilon, \v k) &= i \sigma_z \underbrace{[\vec K \cdot (\sigma_z, \sigma_y, - \sigma_x)^T + i m(K)]}_{\equiv \tilde g(\vec K)},
\end{align}
where 
\begin{equation}
    \mathcal{G}(\vec K) = -\frac{\sqrt{3}}{J} \Upsilon_{r+m}(K)F_{r+m}(K)
\end{equation}
and 
\begin{equation}
    m(K) = \frac{2 m}{F_{r+m}(K)}
\end{equation}
\end{subequations}

Note that for constant $m \neq m(K)$, $\tilde g(\vec K)$ has the same topology as the non-fractionalized counterpart of our theory. 

We next use that for an arbitrary product of GFs

\begin{equation}
\begin{split}
q[\mathbf G] & = q[\mathcal{G} \tilde G] = q[\mathcal G] +q[\tilde G] \\ & + 2\epsilon_{\mu \nu \rho} \partial_\mu \tr [(\mathcal G^{-1} \partial_\nu \mathcal G)(\tilde G^{-1} \partial_\rho \tilde G)] \\ & = q[\tilde G] = q[\tilde g].
\end{split}
\label{eqSub:q}
\end{equation}
Since  $\mathcal G(K)$ is a function instead of a matrix, $q[\mathcal{G}]$ vanishes. The second equality follows because the surface term vanishes under the integral sign. This is shown in the following. We start by noting 
\begin{equation}
    \tr [(\mathcal G^{-1} \partial_\nu \mathcal G)(\tilde G^{-1} \partial_\rho \tilde G)] = (\mathcal G^{-1} \partial_\nu \mathcal G)\tr[\tilde g^{-1} \partial_\rho \tilde g].
    \label{eqSub:trace}
\end{equation}
The factor $(\mathcal G^{-1} \partial_\nu \mathcal G) = (\mathcal G^{-1} \partial_K \mathcal G) \hat K_\nu $ behaves asymptotically at infinity as $1/K$ and is proportional to $\hat{K}=\vec K/K$. We next calculate (here $\tilde{\bm{\sigma}} = (\sigma_z, \sigma_y, - \sigma_x)$, note it has the correct SU(2) algbra)
\begin{equation}
\begin{split}
    \tilde g^{-1}\partial_\rho \tilde g & = \frac{\vec K\cdot \tilde{\bm \sigma} + im(K)}{K^2 + m^2(K)}\partial_\rho [\vec K\cdot\tilde{\bm\sigma} - i m(K)] \\ & = \frac{\vec K\cdot \tilde{\bm\sigma} + im(K)}{K^2 + m^2(K)} [\tilde{\sigma}_\rho - i m'(K)\hat{K}_\rho] \\ & = -i\frac{M_{\rho\mu}}{K^2 + m^2(K)} \tilde{\sigma}_\mu+ h(K)\hat{K}_\rho\mathbf{1},
\end{split}
\end{equation}
where 
\begin{equation}
    M_{\rho\mu} = \epsilon^{\rho\nu\mu}K_{\nu} + m(K)\delta_{\rho\mu} - \frac{K_\rho K_\mu}{K}m'(K),
\end{equation}
and 
\begin{equation}
    h(K) = \frac{K+m(K)m'(K)}{K^2 + m^2(K)}.
\end{equation} 
In this equation, a $'$ indicates a $K$-derivative.  Only those terms proportional to $\mathbf 1$ survive under the trace \eqref{eqSub:trace}. Therefore, both terms $(\mathcal G^{-1} \nabla \mathcal G) \propto \vec K$, and $\tr[\tilde g^{-1} \nabla \tilde g] \propto \vec K$ are proportional to $\vec K$. Therefore, their cross product vanishes which ultimately proves the validity of equation \eqref{eqSub:q}.  

Now, we can calculate $N_3$ according to equation \eqref{eq:VolovikSM}. Note that the part of $\tilde g^{-1} \partial_\mu \tilde g$ which is proportional to $\mathbf 1$ immediately drops out of $q[\tilde g]$. Thus, 

\begin{align}
N_3 &= i\frac{\pi}{3} \int (\text dK) \epsilon_{\mu \nu \lambda} \frac{M_{\mu \rho}M_{\nu \sigma}M_{\lambda \tau}}{(\text K^2 + m^2(K))^3} \tr[\sigma_\rho \sigma_\sigma \sigma_\tau] \notag \\
& = - \frac{2\pi}{3} \int (\text dK) \epsilon_{\mu \nu \lambda} \epsilon_{\rho \sigma \tau}  \frac{M_{\mu \rho}M_{\nu \sigma}M_{\lambda \tau}}{(K^2 + m^2(K))^3} \notag \\
& = - {4\pi} \int (\text dK)  \frac{1}{(K^2 + m^2(K))^3} \notag \\
&\times [M_{xx} M_{yy} M_{zz}+M_{xy} M_{yz} M_{zx}+M_{xz} M_{yx} M_{zy} \notag \\
& - M_{yx} M_{xy} M_{zz}-M_{yy} M_{xz} M_{zx}-M_{yz} M_{xx} M_{zy} ]\notag \\
& =  4\pi \int (\text dK) \frac{m(K) - K m'(K)}{(m(K)^2 + K^2)^2} \notag \\
& = \frac{2}{\pi} \int_0^\infty \text dK K^2 \frac{m(K) - K m'(K)}{(m(K)^2 + K^2)^2} \notag \\
& =   \text{sign}(m(K))\frac{2}{\pi} \int_0^\infty \text dx x^2 \frac{1}{(1 + x^2)^2} \notag \\
& = \text{sign}(m(K))\frac{1}{2}
\end{align}
In the second to last equation, we introduced $x = K/\vert m(K)\vert$ (in our model, $m(K)/m$ is positive definite).
This proofs that $N_3$ is the same for fractionalized and non-fractionalized continuum models. 

\section{Momentum conserving tunneling}
\label{sec:SM:MomCons}

In this section we present details about the momentum conserving tunneling  proposed as a probe of zeros in the discussion section of the main text. 

Consider a tunneling Hamiltonian 
\begin{equation}
H_{\rm tun} = -t \sum_{l \in {t/b}} \int (dk) c^\dagger (\v k) p_{l} (\v k) + H.c.,
\end{equation}
where $p_{l}(\v k)$ describe lead electrons with lead index $l = t/b$ for the top or bottom lead. For simplicity we consider scalar Green's functions in this appendix -- this explicitly includes the case of 1D planar tunneling between quantum anomalous Hall edge states and zeros.

\begin{figure}
    \centering
    \includegraphics{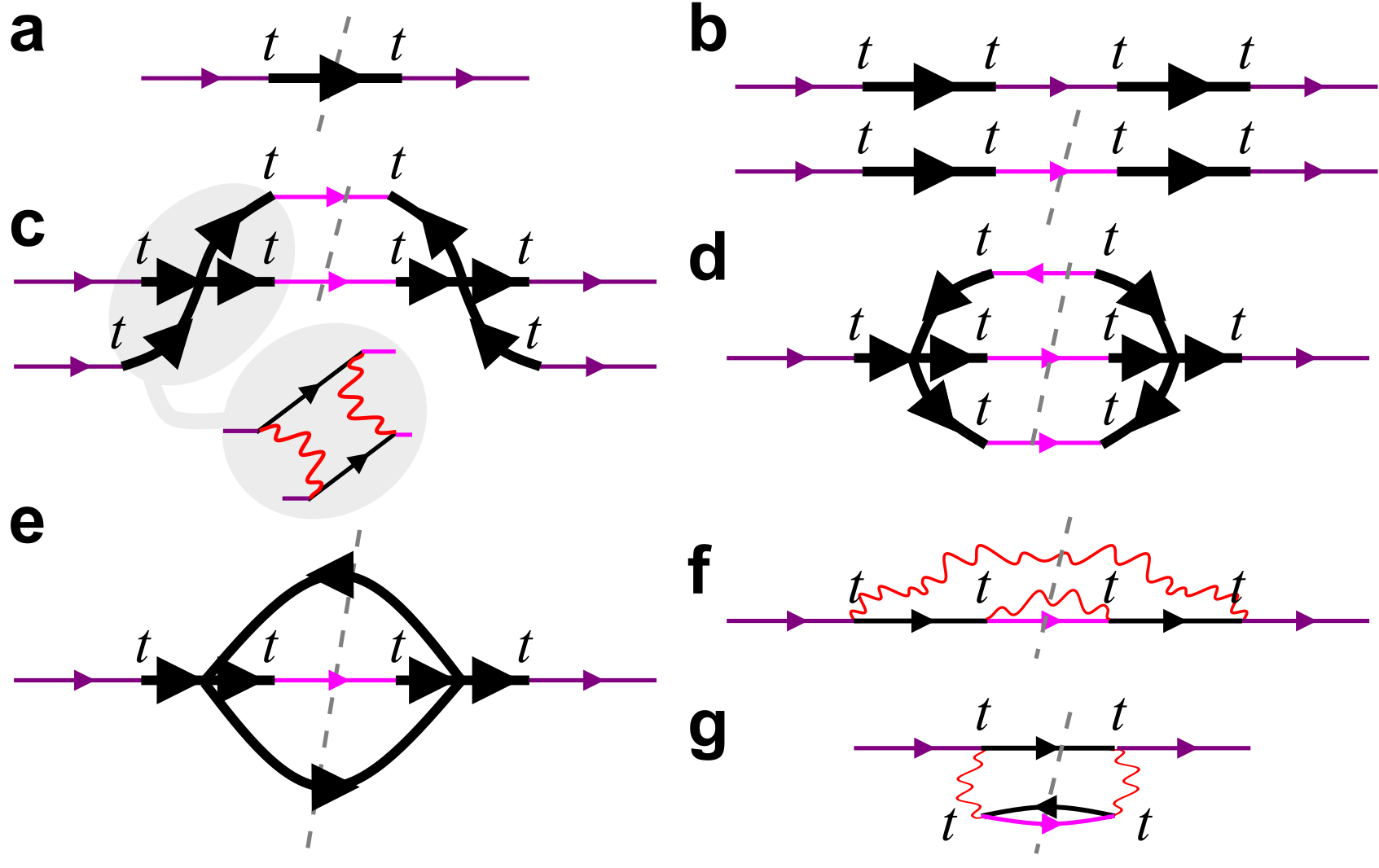}
    \caption{Contributions to the self-energy of the electrons in the upper lead of Fig.~\ref{fig:ContLimit} d). All explanations in the main text of the supplement.}
    \label{fig:Selfenergies}
\end{figure}

We want to determine the transition rate of charge between the two metallic layers. To this purpose, we study the self-energy of electrons and electron-multiplets in the top lead order by order in $t$, see Fig.~\ref{fig:Selfenergies}, in which non-interacting lead electrons from top/bottom lead are represented by purple/pink and the exact Green's function of electrons in the strongly central layer of Fig.~\ref{fig:ContLimit} d) are represented by thick boldface arrows. Most of the discussion in this section is general, i.e. it is independent of the soluble model discussed in the main text, yet the diagrammatic notation is only suggestive: generally the interaction vertices entering Figs.~\ref{fig:Selfenergies} c), d) involve multiple pairs of internal particle-hole lines (not drawn, but implied). As an illustration for a non-trival effective interaction for the model Eq.~\eqref{eq:Htot}, see the inset of Fig.~\ref{fig:Selfenergies} c).

By means of the optical theorem, cuts through the self-energies (illustrated by dashed gray lines in Fig.~\ref{fig:Selfenergies}) can be used to determine the transition rate $\Gamma$ discussed in the main text. 

First, consider the leading order self-energy, i.e. $\mathcal O(t^2)$. Clearly, the imaginary part of this diagram vanishes when there is a spectral gap in the exact fermion Green's function of the interacting system (the transition rate for charge transfer vanishes).

Next, consider the leading diagrams to fourth order in $t$. The upper diagram in Fig.~\ref{fig:Selfenergies} b) is not one-particle reducible -- it is already taken into account by considering the self-energy discussed in panel a). The imaginary part of the lower diagram of Fig.~\ref{fig:Selfenergies} b) describes the rate of tunneling from top metallic lead to bottom metallic lead through virtual excitations of the central correlated system. Since the imaginary part of the Green's function of the central layer vanishes, the transition rate is given by $\Gamma (\epsilon_{\v p}, \v p) = - t^4 [\text{Re} \mathbf G^R(\epsilon_{\v p}^{\rm lead}, \v p)]^2  \text{Im} G^R_{\rm lead}(\epsilon^{\rm lead}_{\v p}, \v p)$, as quoted in the main text. 

Next, we consider inelastic tunneling and co-tunneling, which are of particular interest for the fractionalized theory associated with Eq.~\eqref{eq:Htot}, because certain limits of four-point functions of fermions behave the same as in Fermi liquids. The co-tunneling process Fig.~\ref{fig:Selfenergies} c) describes a charge 2 transfer between the two leads, but is subleading in orders of $t$. The same is true for the exemplary inelastic process creating a particle-hole pair in the lower lead. 

Inelastic processes which are of the same order as the elastic process discussed in Fig.~\ref{fig:Selfenergies} c) involve creating an excitation in the correlated layer. Fig.~\ref{fig:Selfenergies} e) illustrates such a process leaving an electron-hole pair in the correlated layer (this generally can include spin-excitations). Whether such particle-hole pairs are gapped or gapless depends on details of the model.

To illustrate this, the inelastic processes allowed to $\mathcal O(t^4)$ within the model, Eq.~\eqref{eq:Htot}, of this letter are presented in Fig.~\ref{fig:Selfenergies} f) g). The excitation presented in panel f) is gapped: it requires the energy of the incoming particle (the voltage) to be higher than twice the energy gap of e-particles. Such a process is thus not contributing at the lowest energies (of the order of the bands of dispersing Green's function zeros). In contrast, Fig.~\ref{fig:Selfenergies} g) illustrates a process where a particle-hole pair of ``orthogonal fermions'' is left behind, which allows for low energy inelastic charge tunneling.  

In the bulk, for OTI and ZTI phases one may expect that the excitation energy of this particle-hole exceeds the (topological) gap of zeros. This implies that the bulk probe of zeros perisists, i.e. the dip in the tunneling conductance is still clearly observable.

At the boundary ZTI-OTI we expect boundary zeros but no edge states (poles) at the lowest energy. Thus the inelastic processes Fig.~\ref{fig:Selfenergies} f), g) are gapped at lowest energies while boundary zeros show up in momentum-conserving tunneling. In such a situation(Green's function boundary zeros are present, and boundary poles are absent), momentum-conserving tunneling probes of topological edge zeros should work best.

Finally, at the boundary ZTI-OI we expect both zeros and edge states. We expect that inelastic processes creating particle-hole pairs in the edge display characteristic temperature dependence which can be discriminated from the elastic contribution discussed in the main text and leave their study to a separate work.

\end{document}